\documentclass[twocolumn,pre,showpacs,amsfonts,amsmath,assymb,eufrak,longbibliography]{revtex4-2}
\usepackage{grffile}
\usepackage{amsfonts}
 \usepackage{setspace}
\usepackage[utf8]{inputenc}
\usepackage{boondox-cal}
\usepackage{bbm}
\usepackage{epsfig}
\usepackage{color}
\usepackage{bm}
\usepackage{amssymb}
\usepackage{subfigure}
\usepackage{textcomp}
\usepackage[utf8]{inputenc}
\usepackage{textcomp}
\usepackage{amsmath}
\usepackage{esint}
\usepackage{xcolor}
\usepackage{subfigure}
\usepackage{float}
\usepackage{soul}
\usepackage{stackrel}
\usepackage{color}
\usepackage[bottom]{footmisc}
\usepackage[
colorlinks=true, 
pdfstartview=FitV, 
linkcolor=blue, 
citecolor=magenta, 
urlcolor=blue, 
bookmarks=true,
bookmarksnumbered=true,
pdftitle={},
pdfauthor={}
]{hyperref}
\usepackage{epsfig,amsopn}
\usepackage{graphicx}

\definecolor{pink}{rgb}{1,0,0.6}

\newcommand\ra{\rangle}
\newcommand\la{\langle}
\newcommand\nn{\nonumber}
\newcommand\f{\frac}
\newcommand\p{\partial}
\graphicspath{{./Figures/}{../Entropy/}}

\newcommand{\be}{\begin{equation}}
\newcommand{\ee}{\end{equation}}
\newcommand{\bea}{\begin{eqnarray}}
\newcommand{\eea}{\end{eqnarray}}

\def\nn{\nonumber}

\definecolor{dgreen}{rgb}{0,0.7,0}

\def\p{\partial}

\begin{document}

\newcommand{\titlename}{Entropy growth during free expansion of an ideal gas}

\title{\titlename}

\author{  Subhadip Chakraborti$^1$, Abhishek Dhar$^1$, Sheldon Goldstein$^2$, Anupam Kundu$^1$ and Joel L. Lebowitz$^2$}

\affiliation{$^1$International Centre for Theoretical Sciences, Tata Institute of Fundamental Research, Bengaluru 560089, India}%
\affiliation{$^2$Departments of Mathematics and Physics, Hill Center, Rutgers, The State University of New Jersey, 110 Frelinghuysen Road, Piscataway, New Jersey 08854-8019, USA}

\date{\today}
\begin{abstract} 
To illustrate Boltzmann’s construction of an entropy function that is defined for a  microstate of a macroscopic system, we present here the simple example of the free expansion of a one dimensional gas of non-interacting point particles. The construction requires one to define macrostates, corresponding to macroscopic variables. 
We define a macrostate $M$ by specifying the fraction of particles in rectangular boxes $\Delta x \Delta v$ of the single particle position-velocity space $\{x,v\}$. We verify that when the number of particles is large the Boltzmann entropy, $S_B(t)$, of a typical microstate of a nonequilibrium ensemble coincides 
with  the Gibbs entropy of the coarse-grained  time-evolved one-particle distribution associated with this ensemble. $S_B(t)$ approaches its maximum possible value for the dynamical evolution of the given initial state. The rate of approach depends on the size of $\Delta v$ in the definition of the macrostate, going to zero at any fixed time $t$ when $\Delta v \to 0$. Surprisingly the different curves $S_B(t)$ collapse when time is scaled with $\Delta v$ as: $t \sim \tau/\Delta v$. We find an explicit expression for $S_B(\tau)$ in the limit $\Delta v \to 0$. We also consider a different, more hydrodynamical, definition of macrostates for which $S_B(t)$ is monotone increasing, unlike the previous one which has small decaying oscillations near its maximum value. Our system is   non-ergodic, non-chaotic and  non-interacting;  our results thus illustrate that these concepts  are not as relevant as sometimes claimed, for observing macroscopic irreversibility  and entropy increase. Rather, the notions of initial conditions, typicality, large numbers and coarse-graining   are the important factors. We demonstrate these ideas through extensive simulations as well as analytic results.
\keywords{Boltzmann's entropy, Free expansion, Thermalization}
\end{abstract}
\maketitle

\section{Introduction}
\label{Introduction}
\noindent
According to the second law of thermodynamics, any spontaneous change in an isolated system  leads to an increase of the  thermodynamic entropy, $S$ (as defined by Clausius).  The second law thus provides in a sense an arrow of time and quantifies the irreversibility that we observe in everyday physical phenomena. Understanding how such irreversibility emerges from the microscopic reversible Newtonian dynamics of a many-particle system  was the remarkable achievement of Boltzmann. He pointed out the key idea that the observed irreversibility is the typical macroscopic behavior given appropriate initial conditions, that becomes a certainty when we take the system size truly macroscopic. Boltzmann also provided a clear prescription for the construction of an entropy function (which we denote as $S_B$) that is defined for a \emph{single} microstate of a macroscopic system in a given macrostate. This entropy function is  defined  for a system in or out of equilibrium. It is equal  to the thermodynamic entropy for a system in equilibrium.

The deep and somewhat subtle ideas of Boltzmann~\cite{boltzmann1897} have been widely discussed~\cite{feynman2017,lanford1976,penrose89,greene2004} and   clarified in recent work~\cite{Lebowitz_PA1993,Lebowitz_PT1993,Lebowitz_Book2008,Goldstein_Book2020}. We mention here a particularly relevant quote from Ref.~\cite{lebowitz2021}: 
{\it Time-asymmetric behavior as embodied in the second law of thermodynamics is observed in individual macroscopic systems. It can be understood as arising naturally from time-symmetric microscopic laws when account is taken of a) the great disparity between microscopic and macroscopic sizes, b) initial conditions, and c) that what we observe are “typical” behaviors — not all imaginable ones. Common alternate explanations, such as those based on equating irreversible macroscopic behavior with ergodic or mixing properties of ensembles (probability distributions) already present for chaotic dynamical systems having only a few degrees of freedom or on the impossibility of having a truly isolated system, are either unnecessary, misguided or misleading.}

The present work is an attempt to provide a numerical demonstration of some of the above ideas presented in \cite{Goldstein_Book2020} through a simple example.

Our microscopic model is a  gas of $N (\gg 1)$ non-interacting point particles of unit masses confined to move inside a one-dimensional box of length $L$.  Initially the gas is in thermal equilibrium (to be defined more precisely later) and confined, by a partitioning wall, to the left half of the box.  We consider its subsequent evolution on removal of the partition. In our work we consider two distinct (families of)  macroscopic variables. For the first,  we consider a coarse graining of the single particle phase space $\{ \mu \equiv (x,v) \}$ into rectangles $\Delta_\mu$ with volumes $\Delta x \Delta v$ and look at the distribution  $f(x,v,t)$,  at time $t$, of particles in this space. This leads to a definition of $S_B$,  that we refer to as $S_B^f$.  The second macroscopic description
is given by the three locally conserved fields  $U=\{\rho(x,t), p(x,t), e(x,t)\}$  corresponding to mass, momentum and energy --- defined using a spatial coarse-graining.   The Boltzmann entropy corresponding to $U$ will be referred to as $S_B^{U}$. 

 We study the time evolution of the two choices of macrovariables, $f$ and $U$, and the associated entropies, $S_B^f, S_B^{U}$.  The simplicity of the model allows us to perform highly accurate simulations with large number of particles (of order $10^7$) and compute both mean distributions (averaged over initial ensembles) analytically as well as empirical ones (with single realizations). The results from the empirical distributions allow us to test the  typicality of macroscopic behavior, i.e.,  that typical microstates---that is, the overwhelming majority of microstates---corresponding to a given coarse-grained description defining a macrostate in terms of values of certain  macrovariables yield the same future behavior  of those macrovariables, with some tolerance. We assume, as is usually done in statistical mechanics, that ``overwhelming majority"  is defined with respect to  the projection of the microcanonical measure onto the relevant macrostate, i.e. the uniform distribution on the region of phase space corresponding to  the macrostate.

We find that as expected, both $S^f_B$ and $S^U_B$ approach for long times their equilibrium values with the behavior of a typical microstate being the same as that averaged over the initial ensemble. There are however some interesting surprises in the time evolution of $S^f_B(t)$. The rate at which $S^f_B(t)$ increases depends strongly on $\Delta v$ with $d S^f_B(t)/dt$ apparently going to zero as $\Delta v \to 0$. However, upon rescaling time, $t \to \tau/\Delta v$, the different curves  collapse to a single curve $S^f_B(\tau)$. This curve has small decaying oscillations near its maximum. We obtain an analytic expression for $S^f_B(\tau)$ which agrees with the observations. There are no such surprises for $S^U_B(t)$ which increase monotonically to the equilibrium value.

We want to point to some of the earlier studies related to this issue. Of particular relevance are the works of Frisch~\cite{frisch1958approach} and of De Bievre and Parris~\cite{de2017}, who studied the time evolution of an initially spatially non uniform ideal gas, as do we. They do not however consider the entropy change in the process, which is our focus here. The paper ~\cite{de2017} is particularly nice and highly recommended. It explains in a clear and rigorous way the resolution of the objections to Boltzmann by Zermelo and Loschmidt --- i.e., the ``paradoxes" of Poincare recurrence times in a finite system and of reversibility of the microscopic dynamics.

The evolution of  Boltzmann's  entropy  has been earlier  investigated numerically in interacting  systems such as fluid models~\cite{AlderWainwright1956,orban1967,Lavesque_JSP1993,Rochin_JSP1997} and  in systems evolving via maps~\cite{FalcioniPhysicaA2007}.  Some subtleties for dense fluids were pointed out by Jaynes in  \cite{Jaynes_PRA1971}, discussed further in \cite{Goldstein_PD2004} and numerically investigated in \cite{Garrido_PRL2004}. 
The one dimensional gas of equal mass hard point particles and hard rods was extensively studied earlier as one of the tractable models where dynamical properties can be obtained analytically and where the question of entropy increase has been investigated. Some of the interesting questions addressed concern dynamical correlations and the evolution of the single particle distribution function~\cite{frisch1958approach,jepsen1965,lebowitz1967,Lebowitz_PR1968,Percus_POF1969,Roy2013,kundu2016}. The Euler hydrodynamic equations for the hard rod system were first obtained in \cite{Boldrighini_JSP1983} and have more recently been discussed in \cite{Doyon_JSM2017} as an example of an interacting integrable model, where it is also shown that there are dissipative Navier-Stokes corrections which vanish when one goes from rods to point particles.   
The effect of integrability-breaking on entropy  growth was studied in \cite{Cao_PRL2018} for hard rods in a harmonic trap.   Boltzmann's ideas also appear in recent discussions of thermalization in isolated  quantum systems~\cite{tasaki2016,mori2018}.  In contrast to these  studies, the present work considers the case of a completely non-interacting system, namely the ideal gas in one dimension.

The plan of the paper is as follows. In Sec.~\ref{sec:model} we  define the Boltzmann entropy for a general classical macroscopic system. We then describe the precise model and the different  choices of macrostates. In Sec.~\ref{sec:numerics} we present our numerical results on the evolution of the macroscopic fields and the entropy functions for the two different choices of macrovariables.  This section also contains the derivation of the expression for $S^f_B(\tau)$ in the rescaled time $\tau$.  For $S^U_B$, we  present an analysis of the results based on  the  ``hydrodynamic'' equations for the macroscopic fields. In Sec.~\ref{sec:atypical} we study how these macrovariables and the associated (Boltzmann) entropies evolve with time for atypical initial conditions.  A geometric picture of the dynamics in phase space is provided in Sec.~\ref{sec:geometric} and we conclude with a discussion in Sec.~\ref{sec:conclusions}. Some exact results for the evolution of the macroscopic fields  are presented in App.~\ref{sec:appendix-field}.

\section{Boltzmann's entropy, definition  of the microscopic model and choice of macrostates}
\label{sec:model}
\subsection{Boltzmann's entropy}

The microstate of a classical system of $N$ particles of unit mass confined in a box, denoted by $X$, is specified by the positions ${\bf x}_i$ and velocities ${\bf v}_i={\bf p}_i$,  with $i=1,2,...,N$, \emph{i.e.}, $X=({\bf x}_1,{\bf x}_2,\ldots {\bf x}_N,{\bf v}_1,{\bf v}_2,\ldots,{\bf v}_N)$. The dynamics of the system is given by a Hamiltonian $H(X)=\sum_{i=1}^N v_i^2/2$.

We now consider a macroscopic or ``coarse-grained'' description for the case $N\gg 1$. A simple example of such a description is provided by the macrovariable $N_{\rm left}$ which gives the total number of particles in the left half of the box. Clearly, this is a function of the microstate $X$ and we can write $N_{\rm left}(t)=M(X(t))$,  with $M(X) = N_{\rm left}(X)$.

In general we can describe a macrostate by specifying a set of macrovariables $M(X) = \{M_1(X), M_2(X), . . . , M_n(X)\}$, with resolution $\Delta M = \{\Delta M_j \}$~\cite{Goldstein_Book2020}. We identify these macrostates with the elements of a partition of the full phase space $\Gamma$ into sets $\Gamma_{\hat M}$ of the form
\begin{align}
\label{gammaM}
\Gamma_{\hat{M}} = \{ X \in \Gamma | \hat{M}_j \leq M_j(X) \leq \hat{M}_j + \Delta M_j,  j = 1, ... , n\}.
\end{align}
These provide  a coarse-grained description in the sense that many different $X$ correspond to the same range of values of the macrovariable $M(X)$, and hence to the same set $\Gamma_{\hat{M}}$. 

Each microstate $X$ belongs to some set $\Gamma_{\hat{M}}$ corresponding to the coarse-grained value of the macrovariable $\hat{M}=\hat{M}(X)$ (thus for $X\in \Gamma_{\hat{M}}$ as in Eq.~\eqref{gammaM}, $\hat{M}(X) = \hat{M}$).
Boltzmann’s insight was to associate to each microscopic state $X$ an entropy, through the set $\Gamma_{\hat{M}}$ to which it belongs~\cite{Lebowitz_PA1993, Lebowitz_Book2008,Goldstein_Book2020,Goldstein_TM2019}:
\begin{equation}
S_B(X)=S_B[\hat{M}(X)]=\ln |\Gamma_{\hat{M}}|, \label{en-Boltz}
\end{equation}
where we have set Boltzmann's constant $k_B=1$. The volume of the  set  $\Gamma_{\hat{M}}$  is 
\begin{align}
|\Gamma_{\hat{M}}| &= \int \prod_{i=1}^N \frac{d{\bf x}_i d{\bf p}_i}{h}  ~\mathbbm{1} [X \in \Gamma_{\hat{M}}],
\end{align}
where $\mathbbm{1}$ represents the indicator function and $h$ is a constant with the dimension of angular momentum. Here, without loss of generality, we  set $h=1$. As the system evolves under the Hamiltonian dynamics, the microstate is given by $X(t)$ while the   macrovariable  evolves as $M(t)=M(X(t))$. Consequently the corresponding set  $\Gamma_{\hat{M}}(t) = \Gamma_{\hat{M}(X(t))}$ also evolves, thereby specifying the time evolution of the Boltzmann entropy as $S_B(t) = S_B[\Gamma_{\hat{M}}(t)]$. Boltzmann  argued that for  an isolated  system starting from a microstate corresponding to a low entropy $S_{B}(0)$, the system evolves in such a way that  $S_B(t)$ ``typically" increases for macroscopic systems even though the microscopic evolution is completely time-reversal symmetric. (In what follows we shall drop the hats on the macrovariables, slightly abusing notation.)

Among  all possible macrostates of a system there are two very important ones: the equilibrium macrostate $M_{eq}$ and the initial macrostate $M_{\rm ini}$. The entropy of the initial macrostate $S_B(t=0)=S_B(M_{\rm ini})$ is low by assumption. On the other hand, the 
macro-region $\Gamma_{M_{eq}}$ is overwhelmingly large compared to other macro-regions associated to other macrostates. It is so large that it contains most of the phase space volume of $\Gamma_E$, an energy shell, assumed to contain all the macrostates $M$. More precisely, for large $N$,  the ratio of their volumes $|\Gamma_{M_{eq}}|/|\Gamma_E| \approx 1 - e^{-c N}$ where $c$ is a positive constant \cite{Goldstein_TM2019, Goldstein_ADP2017, Goldstein_Book2019}. This property corresponds  to equilibrium  because the system should stay in (or near) $\Gamma_{M_{eq}}$ for long times, consistent with the observed stationarity in thermodynamic equilibrium.

Since $\Gamma_{M_{\rm eq}}$ takes up almost all the volume of $\Gamma_{E}$, when the system starts from a microstate $X$ belonging to a non-equilibrium macrostate $M_{\rm ini}$, so that $|\Gamma_{M_{\rm ini}}| \ll |\Gamma_{M_{\rm eq}}|$, its microscopic dynamics should `typically' take the microstate to regions $\Gamma_M$ of larger phase space volume and thus of larger entropy $S_B$ and eventually to $\Gamma_{M_{\rm eq}}$,  unless the dynamics given by the Hamiltonian  has strong constraints, for example additional conservation laws, or the initial state $X$ is very special \cite{Goldstein_Book2001}. Hence we expect the quantity $S_B(M)$ to increase for the majority (in fact the overwhelming majority) of microstates in $\Gamma_{M_{\rm ini}}$ except for a few whose total volume relative to $|\Gamma_{M_{\rm ini}}|$ goes to zero in the $N \to \infty$ limit. 
Because of this expectation one can make direct connection between $S_B(M_{\rm eq})$ and $S$ (the thermodynamic entropy) in an equilibrium state as suggested by Boltzmann. For an isolated system in equilibrium with energy $E$, in a box $\Lambda$ of  volume $V$ and $N$ particles \cite{Lebowitz_Book2008}
\begin{align}
S(E,V,N) \approx S_B(M_{\rm eq}(E,\Lambda,N)),~~{\rm for~large}~N.
\label{S(EVN)}
\end{align}
Here $M_{\rm eq}(E,\Lambda,N)$ is the equilibrium macrostate of the system. Its volume $|M_{\rm eq}(E,\Lambda,N)|$ depends effectively  only on $V$ for boxes of  reasonable shape. (If, in addition to the energy $E$ and particle number $N$, there are other conserved quantities, as there are for integrable systems such as those we are considering here, these could be considered as on the same footing as $N$ and $E$, with Eq.~\eqref{S(EVN)} modified accordingly. However, we shall not do so here.)

We briefly comment here on why Gibbs' definition of entropy cannot be used in the nonequilibrium situation. We recall first that  the Gibbs entropy of an equilibrium canonical ensemble $\varrho_{\rm eq}$ is defined as  
\begin{align}
S_G[\varrho_{\rm eq}(X)] = -\int \prod_{i=1}^N dx_i d p_i~\varrho_{\rm eq}(X)~\ln \varrho_{\rm eq}(X), \label{en-gibbs}
\end{align}
and this can be identified for macroscopic systems with the thermodynamic entropy $S$. Extending this definition to the non-equilibrium situation described by an evolving ensemble  $\varrho_t(X)$ one  obtains the corresponding Gibbs-Shannon entropy $S_G(t)=S_G[\varrho_t(X)]$. However, we note that the volume preserving  dynamics is described by the Liouville equation
\begin{align}
\p \varrho/ \p t = \{H,\varrho \}. \label{liouv}
\end{align}
This ensures that this entropy does not change with time, i.e., $d S_G(t)/dt=0$.

A general study of the Boltzmann  entropy and the coarse grained Gibbs entropy, related to the $F_\alpha$ in Eq.~\eqref{F_alpha}, is given in Chapters IV and V of Ref.~\cite{penrose2005}. The discussions there are enlightening. Another book with a clear analysis of the issues discussed in this article is by Oono~\cite{oono2017}.

\subsection{Definition of the model and  choices of macrostates}
Our model  consists of $N$ non-interacting point particles of mass $1$ confined  in a one dimensional box of size $L$.  The Hamiltonian of the system $ H = \sum_{j=1}^N v_j^2/2$ consists of only kinetic energy.  In between collisions with the walls (at $x=0,L$), each particle moves at constant velocity. On collisions  with the walls, the velocities are reversed.

We now describe the two families of macrovariables that we will consider in this study.
\begin{figure*}
\begin{center}
\leavevmode
\includegraphics[width=17.5cm,angle=0]{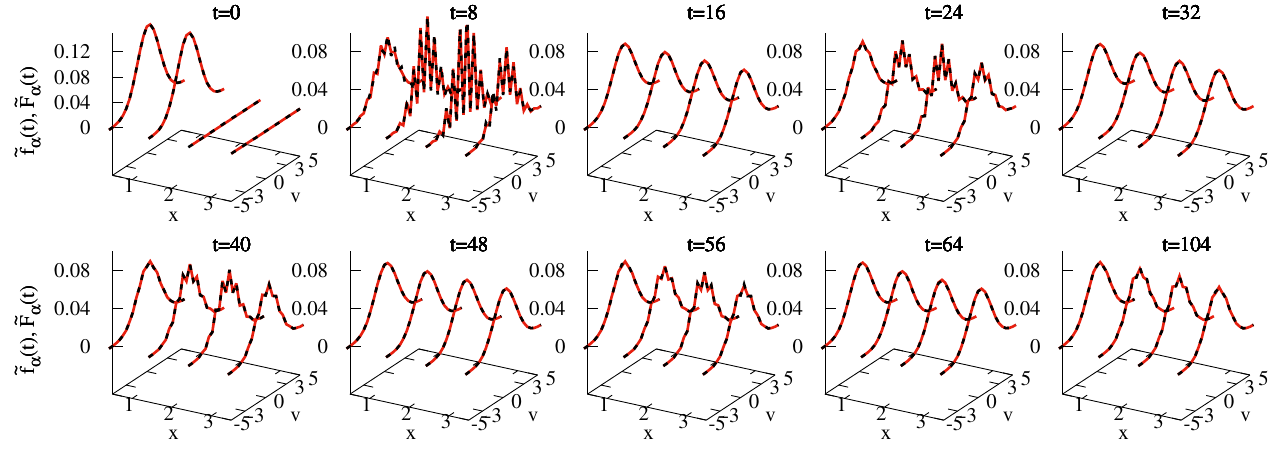}
\caption{Plot of evolution of the empirical  particle density $\tilde{f}_\alpha(t)=f_\alpha(t)/N$ (black dashed lines) starting from a single initial microscopic configuration in the two-dimensional $\mu$-space for grid size $\Delta x=\Delta v=0.5$ and $N=10^7$, $L=4$. In the single initial configuration, the  positions of the particles are distributed uniformly between $(0,L/2)$ and the velocities are drawn from  the Maxwell distribution given by Eq.~\eqref{Maxwell} with canonical temperature $T_0=2.5$. We observe that  $\tilde{f}_\alpha(t)$ approaches towards its equilibrium form at large times, however the convergence is oscillatory as can be seen from the recurrences at times $t=16,~32,~48,~64$ to very close to the equilibrium form. The evolution is also compared with the analytical result for the averaged single particle distribution $\tilde{F}_\alpha=F_\alpha(t)/N$ from Eqs.~\eqref{F_alpha} and (\ref{f_eq2}) which is obtained after averaging over initial configurations chosen from uniform position distribution over $(0,L/2)$ and Maxwell velocity distribution at temperature $T_0=2.5$ (this is the equilibrium state in the left half of the box).
The good agreement between $\tilde{f}_\alpha(t)$ and $\tilde{F}_\alpha(t)$ is a consequence of  typicality.}
\label{fig1}
\end{center}
\end{figure*}

{\bf Choice I --- The distribution of particles in the single-particle phase space}: We consider $\mu$-space $\{(x,v)\}$ and divide it into cells   $\Delta_\alpha$, each of size $| \Delta_\alpha |=\Delta x \Delta v$. For a given microstate $X=\{x_i,v_i\}$ we specify the number of particles $N_\alpha$ in each cell. We then obtain the particle number density in each cell:
\bea
\label{falpha}
f_\alpha= \frac{N_\alpha}{| \Delta_\alpha |}.
 \eea
This satisfies the normalization $\sum_\alpha f_\alpha | \Delta_\alpha |=N$. The set $\{f_\alpha\}$ specifies our  first family of macrovariables, with its corresponding macrostates. The ``number'' of microstates (volume) for a given specification of $\{N_\alpha\}$ is given by $|\Gamma_M|= \prod_\alpha [| \Delta_\alpha |^{N_\alpha}/ N_\alpha !]$~\footnote{ The  phase space volume of $N_\alpha$ identical particles in a box of size $|\Delta_\alpha|$ is given by $\frac{1}{N_\alpha!}(|\Delta_\alpha|/h)^{N_\alpha}$. For the specification $\{ N_\alpha \}$ over the set of all boxes, we then arrive at the product form (with $h$ set to $1$)}. Thus, with  $S^f_B=\ln |\Gamma_M|$, we have using  Stirling's formula for large $N$  the entropy per particle
\bea
s_B^f=S_B^{f}/N=-\frac{1}{N}\sum_\alpha  | \Delta_\alpha | f_\alpha \ln f_\alpha , \label{H-function1}
\eea
up to an additive constant.

To get a handle on the behavior of   $s_B^f$ we also consider, for a finite number of particles, an average (indicated below by $\la...\ra$) over initial microscopic configurations  chosen from a phase space distribution $\varrho_0(\{x_i,v_i\})$:
\bea
\label{single_F}
 F(x,v,t)=  \sum_{i=1}^N \left\la \delta(x_i(t)-x) \delta(v_i(t)-v) \right\ra,
\eea 
where $\{x_i(t),v_i(t)\}$ are the positions and velocities of the particles at time $t$, obtained from the non-interacting dynamics. We note that $F(x,v,t)$ is the single-particle marginal obtained from  the full phase space density $\varrho_t(\{x_i,v_i\})$ with initial distribution $\varrho_0$. For our ideal gas,  $F(x,v,t)$  obeys the autonomous equation 
\begin{equation}
  \partial_t {F}+v \partial_x {F}=0, \label{Feq}
\end{equation}
and can be computed analytically as shown in App.~\ref{sec:appendix-field}.   We can now define a coarse-grained distribution corresponding to a partition of the $\mu$-space as
\bea
F_\alpha = \frac{1}{| \Delta_\alpha |} \int_{x,v \in \Delta_\alpha} dx ~ dv F(x,v,t), \label{F_alpha}
\eea
and a corresponding coarse-grained entropy:
\bea
s^F_\Delta=-\f{1}{N}\sum_\alpha  | \Delta_\alpha | F_\alpha \ln F_\alpha. \label{s^F}
\eea
Note that this has a similar form to Eq.~\eqref{H-function1}; however, here we have used mean distributions instead of the empirical distributions used there.
These will in fact typically be more or less the same, 
\begin{equation}
s^f_B \approx s^F_\Delta,
\end{equation}
a consequence of the law of large numbers crucial for our analysis.

 We note that if we let $|\Delta_\alpha| \to 0$ in Eqs.~\eqref{F_alpha} and \eqref{s^F} then for any fixed $t$, 
\begin{align}
s^F_\Delta \to s^F=-\frac{1}{N}\int dx dv F \ln F,
\end{align}
 which, since the evolution of $F$ satisfies Eq.~\eqref{Feq},  makes $s^F$ independent of $t$ {\footnotetext{It should be noted however that in the limit as $t \to \infty$,  $F(t)$ weakly approaches a uniform spatial distribution (Prosser mixing~\cite{prosser1969})}}.  This would
also be the case for $s^f_B$ when $\Delta_\alpha \to 0$,  $N \to \infty$, and $f_\alpha$ is suitably  normalized. This is due to the fact
that we are dealing with a non-interacting system so that $s^F$ is just, up to normalization, the Gibbs entropy of the entire $N$-particle system (whose distribution can be taken to be the evolving product of $F$'s), which does not change under the time evolution. We shall see later that even
for the ideal gas if we look on a  time scale proportional to $1/\Delta v$ we will
  see $s^F_\Delta \approx s^f_B$ increase with time albeit non-monotonically.

{\bf Choice II - The spatial distribution of mass, momentum and energy}: We divide the box $(0,L)$ into $K$ cells $\delta_a$, $a=1,2...K$, each of size $\ell = L/K$. For a given microscopic configuration $X$, let $N_a$ be the number of particles in cell $\delta_a$ and let $P_a$ and  $E_a$ be the total momentum and total energy of these particles. In this case the macrostate is defined by these  set of locally conserved quantities $U=\{N_a,P_a,E_a\}$ and we obtain the  Boltzmann entropy $S_B^{U}=\ln |\Gamma_U|$ where $|\Gamma_U|$ is the volume of the phase space region $\Gamma_U$ corresponding to the macrostate $U$.  For large $N$ this entropy per particle attains the form
\bea
\label{Boltzmann_entropy2}
s_B^{U}=\frac{S_B^{U}}{N}= \frac{1}{N}\sum_a \rho_a \ell  ~s(\rho_a,\epsilon_a),
\eea
where $s(\rho_a,\epsilon_a)$ is the equilibrium ideal gas entropy per  particle for density $\rho_a=N_a/\ell$ and internal energy density $\epsilon_a=[E_a-P_a^2/(2N_a)]/\ell =e_a-p_a^2/(2\rho_a)$, with $p_a=P_a/\ell, e_a=E_a/\ell$ being the momentum density and total energy density respectively. This is given explicitly (up to additive constant terms) by:
\bea
\label{Boltzmann_entropy3}
s(\rho,\epsilon)= -\ln \rho +  \frac{1}{2} \ln \left(\frac{\epsilon}{\rho} \right)~.
\eea

\section{Results for the time evolution of macrostates and entropy increase}
\label{sec:numerics}

\subsection{Choice I of the macrovariables}
\label{sec:choiceI}

\subsubsection{Numerical results}

We consider $N=10^7$ particles initially uniformly distributed in the left half $(0,L/2)$ of the box with box size $L=4$. For our non-interacting point particle system, the choice of system size $L$ is inconsequential and hence we arbitrarily set $L=4$. Since we keep the system length fixed, changing $N$ corresponds to changing the density in our system. There is no upper bound to the density since there is no interaction between the particles. In real systems the number of particles would scale with the volume.  We consider first the case where the initial velocities of our microstate are drawn from the Maxwell distribution given by \cite{Goldstein_PD2004}
\be
\label{Maxwell}
g_{\rm eq}(v,T_0)=\left( \frac{1}{2\pi  T_0} \right)^{1/2} \exp \left(-\frac{v^2}{2T_0} \right),
\ee
with  temperature $T_0=2.5$.  This is  the canonical ensemble corresponding to the equilibrium macrostate with particles in the left half of the box. We choose a  \emph{single random realization} from this canonical ensemble as our initial microstate. Equivalently we can choose the initial configuration from a microcanonical ensemble with total energy  $E=N T_0/2$. The region $\Gamma_{M_{\rm ini}}$ then consists of all $X \in \Gamma_E$ such that $N_{\rm left}(X)=N$.
 
We divide the $\mu$-space ($x$-$v$ space) of the system into grids of size $\Delta_\alpha=\Delta x \Delta v$ and calculate the evolution of the empirical single particle density $f_\alpha$ given by Eq.~\eqref{falpha} by performing microscopic simulations of the evolution of the given microstate. In Fig.~\eqref{fig1} we plot $\tilde{f}_\alpha=f_\alpha(t)/N$ at different points $x_\alpha,v_\alpha$ in $\mu$-space and  at different times, for $\Delta x= \Delta v=0.5$. We observe that $\tilde{f}_\alpha(t)$ approaches its equilibrium form non-monotonically in time with near-recurrences to the equilibrium distribution.  At large times, the $\tilde{f}_\alpha(t)$ finally reaches the equilibrium form where particles are uniformly distributed between $(0,L)$  and velocities are  Maxwellian  with temperature $T_0=2.5$.
We also compare the empirical $\tilde{f}_\alpha(t)$ (black dashed lines) and mean distribution $\tilde{F}_\alpha(t)=F_\alpha(t)/N$ (red solid lines), calculated for the same grid size, at different times. The mean distribution ${F}_\alpha(t)$ is computed analytically from Eqs.~\eqref{F_alpha} and (\ref{f_eq2}). We find  good agreement between the empirical density $\tilde{f}_\alpha(t)$ and the mean distribution $\tilde{F}_\alpha(t)$ --- a consequence of  the typicality implied by the law of large numbers for  this non-interacting model. We have verified that this agreement is also valid  when we choose the initial random configuration from a microcanonical distribution with energy per particle given by $T_0/2$.

\begin{figure}
\begin{center}
\leavevmode
\includegraphics[width=8.5cm,angle=0]{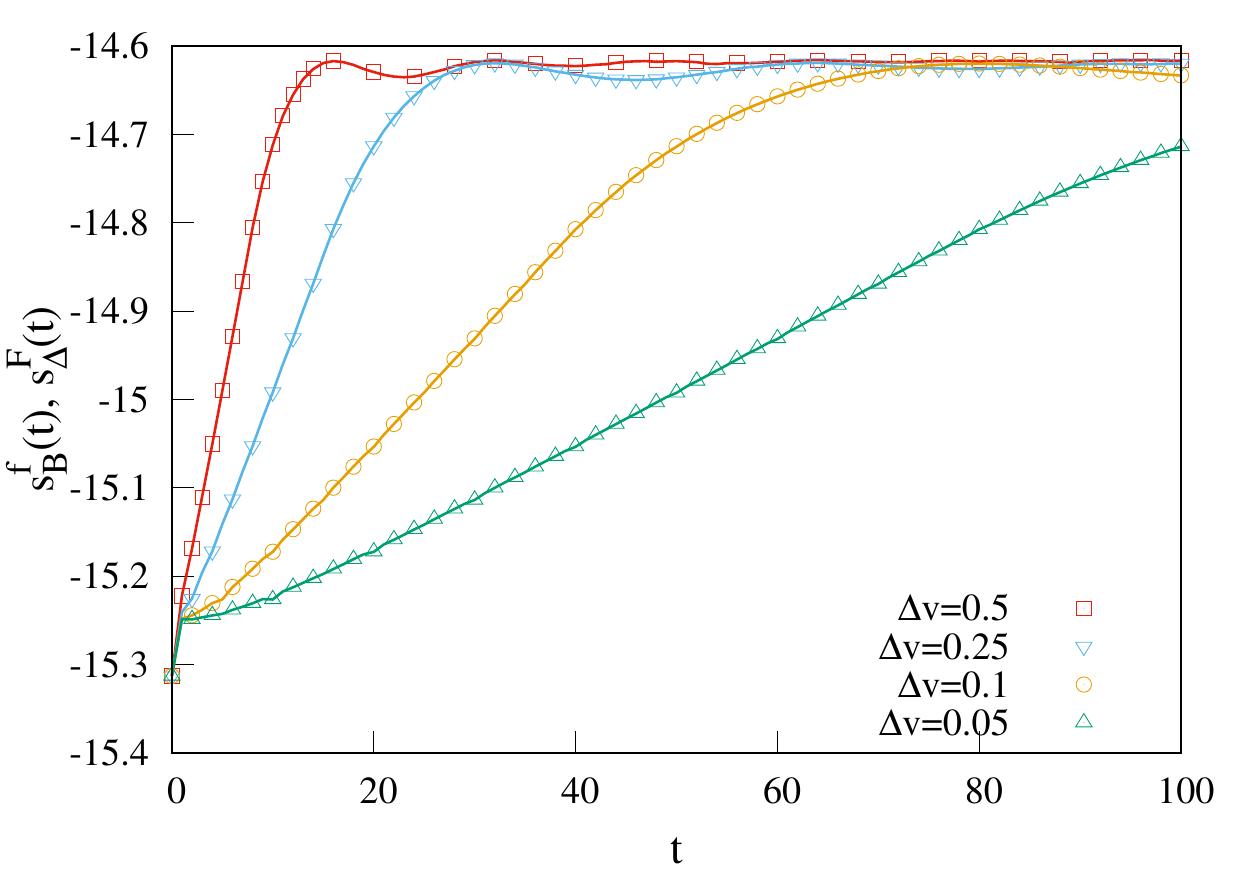}
\caption{A comparison between $s^f_B(t)$ obtained from simulation of a single realization, and $s^F_\Delta(t)$ obtained after an ensemble average, plotted as a function of time during free expansion. For evaluating $s^f_B(t)$ we use the same single initial condition used in Fig.~\eqref{fig1} for $N=10^7$ in a box of size $L=4$. We compute $s^f_B(t)$ for different grid sizes $| \Delta_\alpha |=\Delta x \Delta v$ by keeping $\Delta x=0.5$ fixed and varying $\Delta v=0.5$ (red empty squares), 0.25 (blue empty inverted triangles), 0.1 (yellow empty circles), and 0.05 (green empty triangles). 
The solid lines are $s^F_\Delta(t)$ obtained analytically from Eq.~\eqref{s^F} for $T_0=2.5$. We again observe excellent agreement between  $s^f_B(t)$  and $s^F_\Delta(t)$  and notice that they both eventually increase and saturate to the equilibrium value. However the approach to this equilibrium value is oscillatory  with decaying amplitude and period $2L/\Delta v$.
Also note that  the growth rate at any given time decreases with decreasing  $\Delta v$. }
\label{fig2}
\end{center}
\end{figure}

\begin{figure}
\begin{center}
\leavevmode
\includegraphics[width=8.5cm,angle=0]{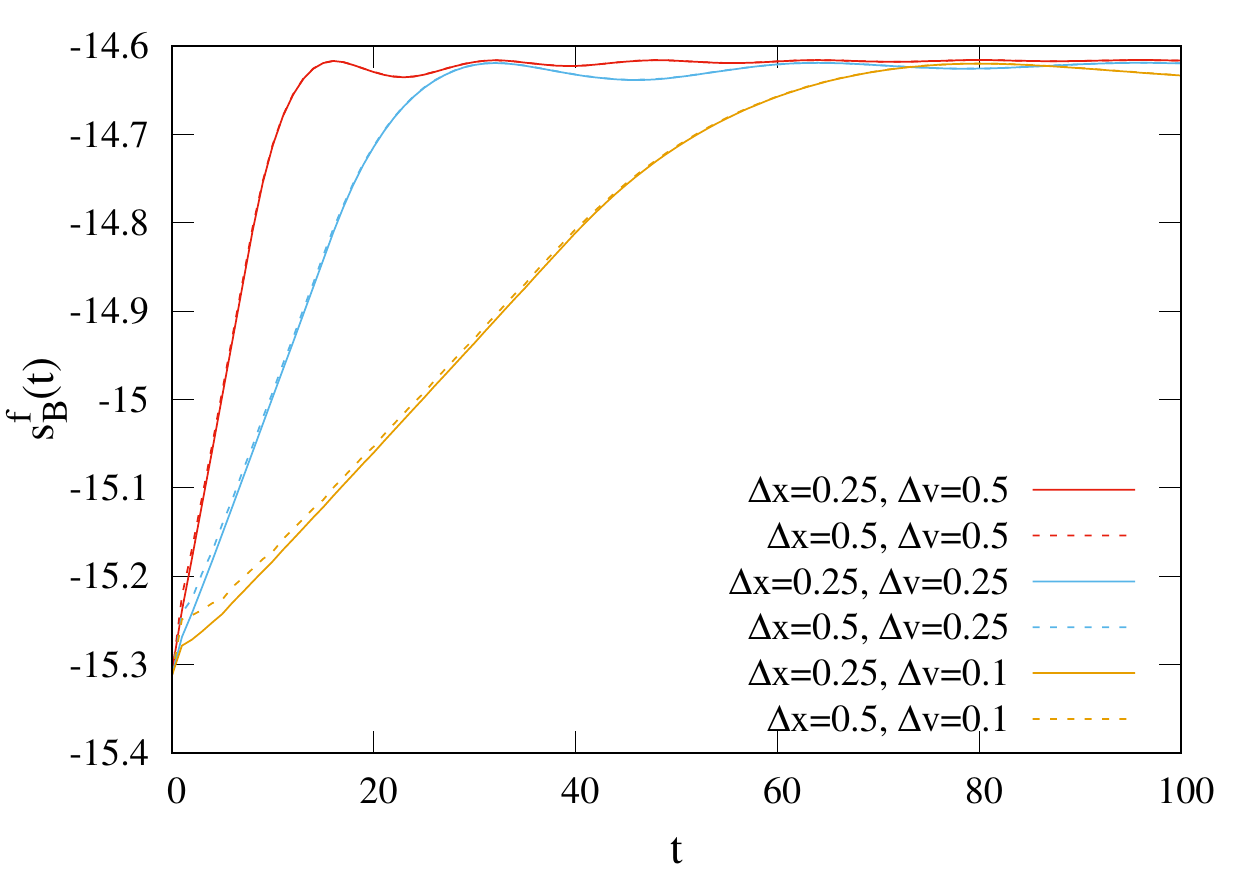}
\caption{Plot showing the dependence of growth of the entropy, $s^f_B(t)$, for different spatial resolutions $\Delta x$. For any given $\Delta v$ we find that there is a weak dependence on the size of $\Delta x$ at early times. The parameter values and initial conditions are the same as in Fig.~\eqref{fig1}.}
\label{fig-delx}
\end{center}
\end{figure}

\begin{figure*}
\begin{center}
\leavevmode
\includegraphics[width=15cm,angle=0]{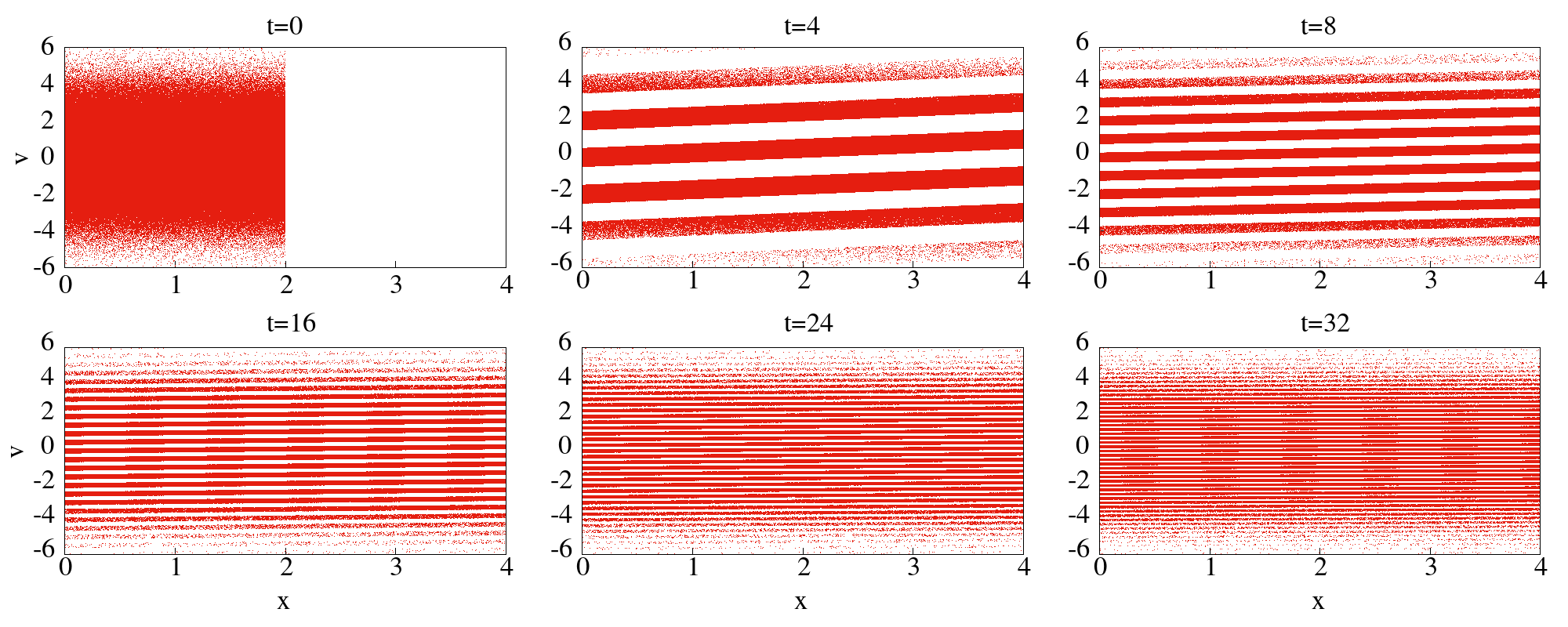}
\caption{Plot showing the distribution of  $N=10^6$ particles in the $(x,v)$ plane at different instants of time. These results correspond to a single realization of the time-evolution of the system, for the same parameters and initial conditions as used in Fig.~\eqref{fig1}. }
\label{fig-muspace}
\end{center}
\end{figure*}

In Fig.~\eqref{fig2} we show the evolution of the corresponding entropy $s^f_B(t)$ [given by Eq.~\eqref{H-function1}, where we fix the additive constant so that at $t=0$, this agrees with Eqs.~(\ref{Boltzmann_entropy2},\ref{Boltzmann_entropy3})] during free expansion,  for the same random single realization  and parameters as  in Fig.~\eqref{fig1}. We plot   $s^f_B(t)$  for different grid sizes  by keeping $\Delta x$ fixed and varying $\Delta v$. The solid lines correspond to the entropy $s^F_\Delta(t)$ [given by Eq.~\eqref{s^F}] calculated from the exact expression for the mean distribution $F(x,v,t)$.  We observe that there is very good agreement between  $s_B^f$ and $s^F_\Delta$, as expected.  Both the entropies grow, initially monotonically with time, touching a value slightly above the final equilibrium value and then   exhibit small oscillations in time with a period $\tau_p=2L/\Delta v$; these eventually die and the entropy saturates to its equilibrium value.  Note that these oscillations were also  seen in the recurrences in Fig.~\eqref{fig1} and we will discuss their origin in the next subsection.  

Though the final increase of entropy appears to be always equal to $\ln (2)$,  we observe in Fig.~\eqref{fig2} that the entropy growth rate decreases with decreasing $\Delta v$. In other words, at any fixed time,  with decreasing $\Delta v$ one observes a correspondingly lower entropy.  On the other hand we see in Fig.~\eqref{fig-delx} that the entropy growth rate shows convergence on decreasing $\Delta x$. This can be understood from the plot of the $\mu$-space distribution shown in Fig.~\eqref{fig-muspace}. We see that with time, the system keeps developing more and more structure in the velocity direction, while, in the spatial direction it becomes more or less homogeneous after some initial time. Thus, decreasing the grid size $\Delta x$ does not give us more information about the system, while decreasing  $\Delta v$ does.

{ To understand the dependence of $s_B^f$ on $\Delta v$ consider the limit of vanishing grid size. For large $N$, corresponding to $f_\alpha$ defined for a given microstate $X=\{x_i,v_i\}$, one can define a smooth function $f(x,v,t)$ such that $N_\alpha = \int_ {x,v \in \Delta_\alpha}dxdv f(x,v,t)$ and $ \int dxdv f(x,v,t)=N$.  Eq.~\eqref{H-function1} then becomes 
 \bea
 {s}_B^{f}(t) \approx -\frac{1}{N}\int dx~dv~ f(x,v,t) \ln f(x,v,t), \label{H-function2}
 \eea
 up to an additive constant.  In the large $N$, small grid size limit,  the function $\tilde{f} = \lim_{\Delta x  \to 0, \Delta v \to 0} \lim_{N\to\infty} f/N$  satisfies the equation 
\begin{equation}
  \partial_t \tilde{f}+v \partial_x \tilde{f}=0, \label{feq}
\end{equation}
using which it follows that its associated ``entropy'' 
 \bea
 \mathcal{s}_B^{f}(t) = -\int dx~dv~ \tilde{f}(x,v,t) \ln \tilde{f}(x,v,t), \label{H-function3}
 \eea
obeys  $d\mathcal{s}_B^{f}(t)/dt=0$.  Thus it would seem that there is no entropy increase in the  large $N$, $\Delta_\alpha \to 0$ limit.  However, as is apparent from  the numerical findings in Sec.~\ref{sec:numerics}, for any fixed grid size $|\Delta_\alpha|$ the exact $s_B^f$, or its approximation on the right hand side of Eq.~\eqref{H-function1}, will typically increase (if initially its value is not at its maximum) over time. For  $N$ large and $| \Delta_\alpha |$ small, significant increase may not begin for a very long time (the time at which $\tilde{f}(x,v,t)$ develops structure on the scale $| \Delta_\alpha |$), a reflection of the fact that the  entropy in Eq.~\eqref{H-function3} does not change with time. }

 The situation is different for a gas of hard spheres  of diameter $a$ in 3D where in the Boltzmann-Grad limit, $a \to 0, N \to \infty$ with $N a^2 \to b>0$, one can define the macrostate by
  a smooth one-particle empirical density which
satisfies the Boltzmann equation, given by  Eq.~\eqref{feq}  modified by collision terms  on the right~\cite{lanford1976,cercignani2013}.   As shown by Boltzmann's $H$-theorem, this  leads to increase of the entropy  $\mathcal{s}^f_B$.

\begin{figure}
\begin{center}
\leavevmode
\includegraphics[width=8.5cm,angle=0]{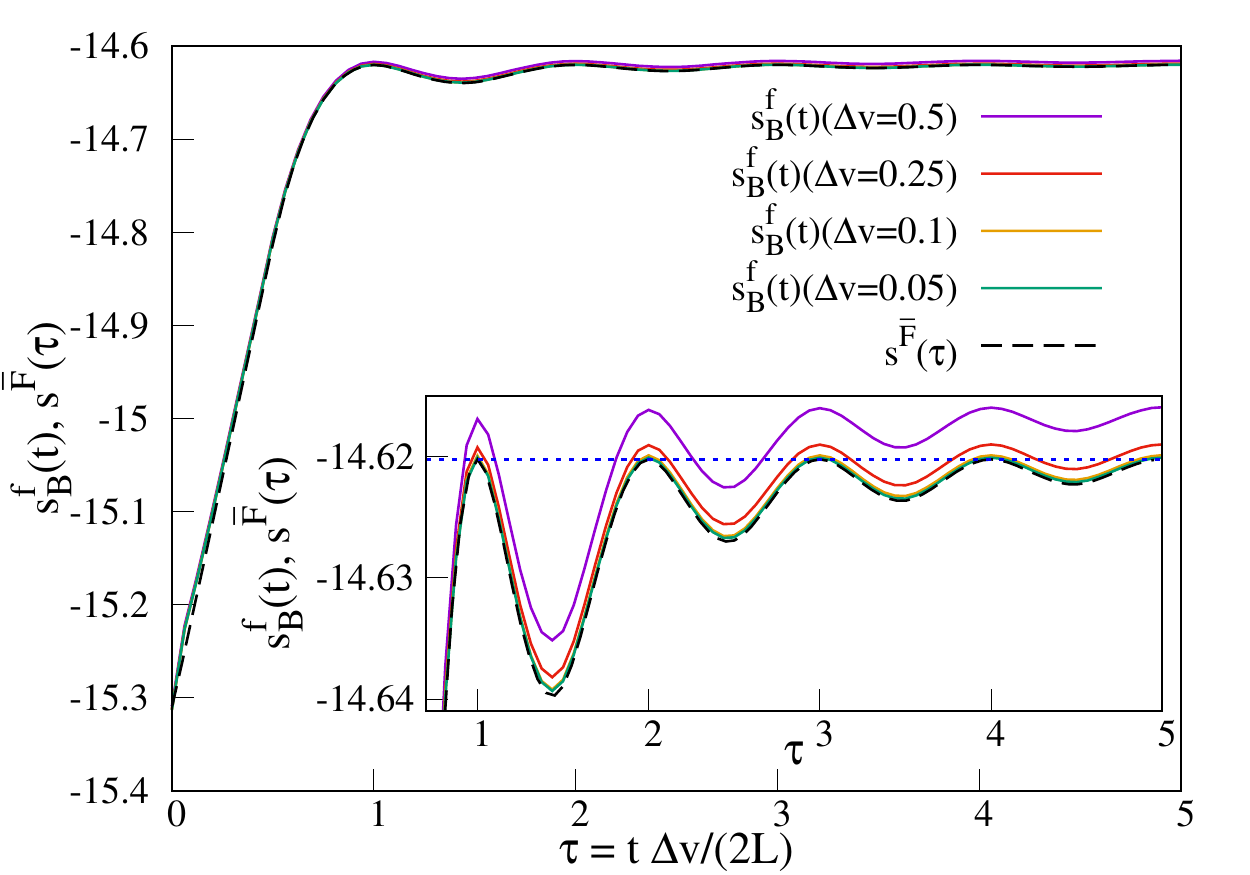}
\caption{  This figure shows a collapse of the  data presented in Fig.~\eqref{fig2} for different values of $\Delta v$, on plotting the entropy as a function of  the scaled time $\tau=t \Delta v/(2L)$. The dashed curve is the analytic prediction from Eq.~\eqref{sbarF} Inset: A zoom up of the plot. The horizontal dotted line corresponds to the entropy given by the right hand side of Eq.~\eqref{H-function2}  with $f = \rho_0 g_{\rm eq}$, where $\rho_0 = N/L$ and $g_{\rm eq}$ is in Eq.~\eqref{Maxwell}. The small overshoot (of $s^f_B$) that we see for larger $\Delta v$ arises since a  coarse-grained velocity distribution is effectively broader than the  Maxwellian from which it comes, leading to a  larger effective temperature.}
\label{fig2scal}
\end{center}
\end{figure}

\subsubsection{Scaling analysis for the evolution of entropy}
\label{app:scaling}

\begin{figure*}
\begin{center}
\leavevmode
\includegraphics[width=15.cm,angle=0]{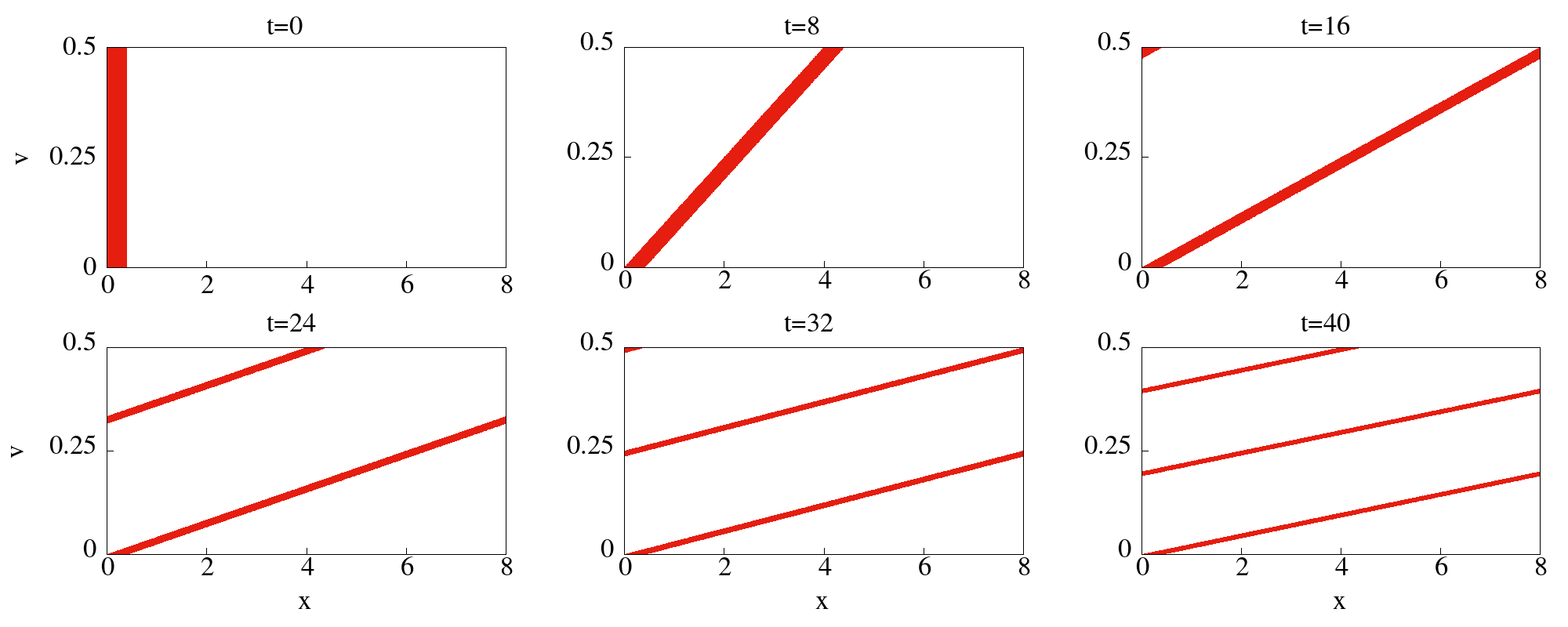}
\caption{Plot showing the evolution of  $N=10^5$ particles, in the $(x,v)$ plane, where the particles move on a circle of length $2L=8$. The particles were initially distributed uniformly in a small box with $ \Delta x =0.25, \Delta v =0.5$. With time the box gets continually stretched and, at times that are multiples of $2L/\Delta v=16$, the stretched pieces wind completely around the box. Comparing with Fig.~\eqref{fig2} we see that the dips in $s^f_B(t)$ occur at times $\approx 24, 40$, at which the winding around the length $2L$ is complete.}
\label{schem}
\end{center}
\end{figure*}

  We now return to the question of the observed oscillation period $\tau_p=2L/\Delta v$ 
in Fig.~\eqref{fig2}. This is easiest to understand once we consider a mapping of the dynamics of particles in a box of length $L$ to the dynamics on a circle of length $2L$. This mapping corresponds to the $\mu$-space map $\phi$ taking $[0,L] \times \mathbb{R}$ to $[0, 2L] \times [0,\infty)$ and given by $(x,v) \to (x,v)$ for $v > 0$ and $(x,v) \to (2L - x, -v)$ for $v < 0$ . With this mapping we see that  two particles, initially at the same spatial point but with  velocity difference $\Delta v$ (size of the velocity grid), will meet again (possibly in different locations) at times $n \tau_p$, where $n$ is an integer.  Furthermore, at the times $t=n \tau_p$, all points that are initially within  a cell $\Delta x \Delta v$  will lie on a narrow strip that winds around precisely $n$ times around the circle and still within $\Delta v$ [see Fig.~\eqref{schem}]. Thus, the spatial distribution of points becomes exactly uniform within the region $v$ to $(v+\Delta v)$ at the times $t= n \tau_p$ and this explains the fact that $s^F_\Delta$ (in Eq.~\eqref{s^F}) reaches its maximum value at these times. At intermediate times, the  winding on the circle is incomplete and we get a lower entropy. 

In Fig.~\eqref{fig2scal}  we show plots of the entropy time-evolution data for different values of $\Delta v$ [from Fig.~\eqref{fig2}] as a function of the scaled time $\tau= t \Delta v/(2L) \equiv t/\tau_p$. We find a remarkable collapse of the data to a single curve. The  physical picture in the preceding paragraph in fact leads to an analytical understanding of this and we can obtain an explicit expression for the evolution of the entropy in the scaled time variable --- this is  given by the function: 
\begin{align}
s^{\bar{F}}(\tau)= -\ln \rho_0+ \frac{1}{2} \ln \f{T}{2} -  \f{1}{2}\int_0^{2} dz R(z,\tau) \ln R(z,\tau),\label{sbarF} 
\end{align}
where $\rho_0 = N/L$ and  $R(z,\tau)$ is known explicitly (see below); the integral above can be numerically computed.  In Fig.~\eqref{fig2scal} we find excellent agreement between the collapsed data and the analytic result.

We now present the details of our analytic understanding of the observed scaling and of Eq.~\eqref{sbarF}. For this we use the mapping  between the dynamics with reflecting boundary conditions and the dynamics in a periodic box. The dynamics on the circle simply consists of rotations at constant positive velocities which implies $F(x,v,t)= F(x-vt,v,0)$ [with $F$ periodic in $x$ with period $2L$].  
We claim that the following  space averaged distribution function  will in fact capture the evolution of the entropy of the system at the rescaled time $\tau=t \Delta v/(2L)$: 
\begin{align}
\label{f_avg1}
\bar{F}(x,v,\tau)&=\frac{1}{2 L \tau} \int_0^{2 L \tau } dx^\prime F(x-x',v,0). 
\end{align}
We define a corresponding Gibbs entropy per particle:
\begin{equation}
\label{def:SbarF}
s^{\bar{F}}(\tau) = -\frac{1}{N} \int_0^{2L} dx \int_{0}^\infty dv~ \bar{F} \ln \bar{F}. 
\end{equation}
More precisely we now show  that in the limit $\Delta x \to 0, \Delta v \to 0$, we get 
\begin{align}
\lim_{\Delta \to 0} s^F_\Delta(2 L \tau/ \Delta v)=s^{\bar{F}}(\tau),
\end{align}
which explains the observed scaling. With the system defined on the circle let us consider the averaged distributions:
\begin{align}
&F_\alpha(\Delta x, \Delta v) \nonumber \\
&= \frac{1}{\Delta x \Delta v} \int_{x_\alpha}^{x_\alpha+\Delta x}  dx'\int_{v_\alpha}^{v_\alpha+ \Delta v} dv' F(x',v',t) \\
&= \frac{1}{\Delta x \Delta v} \int_{x_\alpha}^{x_\alpha+\Delta x}  dx'\int_{v_\alpha}^{v_\alpha+ \Delta v} dv' F(x'-v't,v',0) \nonumber \\
&\approx \frac{1}{\Delta v} \int_{v_\alpha}^{v_\alpha+ \Delta v} dv' F(x_\alpha-v't,v_\alpha,0) \nonumber \\
&= \frac{1}{t\Delta v}  \int_0^{ t\Delta v} dx' F(x_\alpha-v_\alpha t-x',v_\alpha,0) \\
&= \bar{F}(x_\alpha-v_\alpha t,v_\alpha,t;\tau).
\end{align} 
Since $x$ is on the circle, we have for small $\Delta x$ and small $\Delta v$,
\begin{align}
&s^F_\Delta (t)=-\frac{1}{N}  \sum_{\alpha} |\Delta_\alpha| \bar{F}(x_\alpha-v_\alpha t,v_\alpha,\tau) \ln \bar{F}(x_\alpha-v_\alpha t,v_\alpha,\tau) \nonumber \\
&\approx -\frac{1}{N}\sum_\alpha \int_0^{2L} dx \Delta v \bar{F}(x-v_\alpha t,v_\alpha,\tau) \ln \bar{F}(x-v_\alpha t,v_\alpha,\tau) \nn \\
&=-\frac{1}{N}\sum_\alpha \int_0^{2L} dx \Delta v \bar{F}(x,v_\alpha,\tau) \ln \bar{F}(x,v_\alpha,\tau) \nn \\
&\approx -\frac{1}{N} \int_0^{2L} dx \int_0^\infty dv \bar{F}(x,v,\tau) \ln \bar{F}(x,v,\tau) \nn \\
&= s^{\bar{F}}(\tau), 
\end{align}
where  we used the translational invariance in going from the second to the third step.  We note that this result explains the main observations in Fig.~\eqref{fig2scal}, namely slowly decaying oscillatory approach to the final value, with precise returns at integer values of $\tau$.  Since $F$ has period $2L$, we can write it in the form $F(x,v,0)=\rho_0 \bar{g}(v) + \phi(x,v,0)$ where $\rho_0$ is the average density in the original box, $\bar{g}(v) = \rho_0^{-1} \int_0^{2L} dx F(x,v,0)/(2L)$ is the averaged global velocity distribution  [$\bar{g}$ is normalized to $1/2$ on $v > 0$] while $\phi(x,v,0)$ has period $2L$ and mean zero (i.e. $\int_0^{2L} dx  \phi(x,v,0)=0$). Hence clearly $\psi(x,v,\tau) = \int_0^{2 L \tau} dx' \phi(x-x',v,0)/(2L)$ is periodic in $\tau$ with period $1$. Then  we have from  Eq.~\eqref{f_avg1} that 
\begin{align}
\bar F(x,v,\tau) =  \rho_0 \bar{g}(v)+ \frac{\psi(x,v,\tau)}{\tau}. \label{fatn1}
\end{align}
We see that for integer values of $\tau = 1, 2, 3, \ldots$,  
$\bar F$ attains the value  $\rho_0 \bar{g}(v)$, which yields the  time-maximum of the entropy $s^{\bar{F}}$ given by:
\begin{align}
s^{\bar{F}}_{\rm max}&=-\f{1}{N}\int_0^{2L} dx \int_0^\infty dv \rho_0 \bar{g}(v) \ln [\rho_0 \bar{g}(v)] \nn  \\
&=-\ln \rho_0 - 2\int_0^\infty dv \bar{g}(v) \ln [\bar{g}(v)].\label{smax}
\end{align}
We also see that the deviations of $\bar{F}$, from the value  $\rho_0 \bar{g}(v)$,  that occur at values of $\tau$ between these integers are at most of order $1/\tau$, implying that the same thing is true for the entropy.

For the special initial condition $F(x,v,0)=\rho_0[1+a \cos(\pi x/L)] h(v) \Theta(v)$, with $h(v)$ any even velocity distribution, we get $\bar{F}(x,v,\tau)=\rho_0\{1 -a [\sin [\pi (x/L-2\tau)]-\sin [\pi x/L]]/(2 \pi \tau) \} h(v)$. We thus explicitly find here that $\bar{F} = \rho_0 h(v)$ for $\tau = 1,2,3, ...$, with  $1/\tau$ deviations for intermediate values as described above. In particular $\bar{F}$ approaches $\rho_0 h(v)$ as $\tau \to \infty$.

Next, we consider the case discussed in Sec.~\eqref{sec:choiceI} where the gas is initially  confined on the left half of the box.  The initial distribution considered is of the product form $F(x,v,0)=\rho(x) g_{\rm eq}(v) \Theta(v)$ and we then get:
\begin{align}
\bar{F}(x,v,\tau) &=g_{\rm eq}(v) \Theta(v) \frac{1}{2 L \tau} \int_0^{2 L \tau} dx^\prime \rho_c(x-x'),
\end{align}
where $\rho_c(x)$ is the initial density profile on the circle $[0,2L]$, given by  $\rho_c(x)=2 \rho_0$ for $x \in [0,L/2] \cup [3L/2,2L]$ and zero elsewhere. To perform the above integral, we Fourier-decompose the density profile as
$\rho_c(x)= \sum_{n=-\infty}^\infty a_n e^{i n \pi x/L}$, where $a_n=(2L)^{-1}\int_0^{2L} \rho_c(x) e^{-i n \pi x/L}$. Hence we obtain:  
\begin{align}
\bar{F}(x,v,\tau)&= g_{\rm eq}(v) \Theta(v) \left[ {\rho_0} + \sum_{n\neq0} \frac{a_n e^{i n \pi (x/L-\tau)} \sin  (n \pi \tau)}{n \pi \tau} \right].
\end{align}
For our initial condition with a half-filled box one finds $a_n=2 \rho_0 \sin (n \pi/2)/(n \pi)$ for $n \neq 0$. Hence we get
\begin{align}
&\bar{F}(x,v,\tau)= g_{\rm eq}(v) \Theta(v) \rho_0 R(x/L,\tau),~~{\rm where}  \\
&R(z,\tau) \nn \\
& = \left[1+ \frac{4}{\pi^2 \tau} \sum_{n=1}^\infty \frac{\cos [n \pi (z-\tau)] \sin (n \pi/2) \sin ( n \pi \tau)}{n^2}\right].
\end{align} 
The product form of $\bar{F}$ leads to simplifications  for the entropy given by Eq.~\eqref{def:SbarF}  and, we finally obtain Eq.~\eqref{sbarF} [after fixing additive constants so that at $t=0$, it agrees with Eqs.~(\ref{Boltzmann_entropy2},\ref{Boltzmann_entropy3})].

\begin{figure*}
\begin{center}
\leavevmode
\includegraphics[width=5.8cm,angle=0]{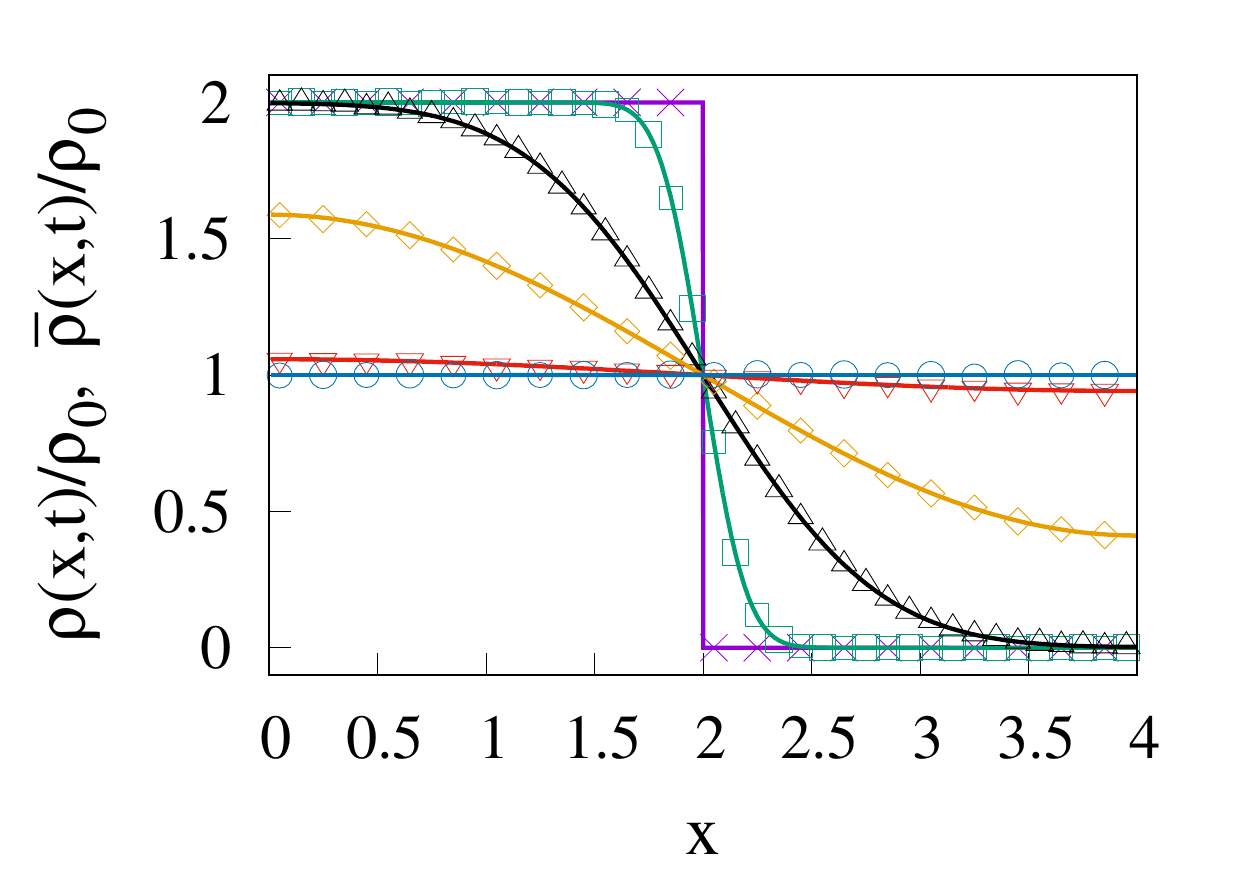}
\includegraphics[width=5.8cm,angle=0]{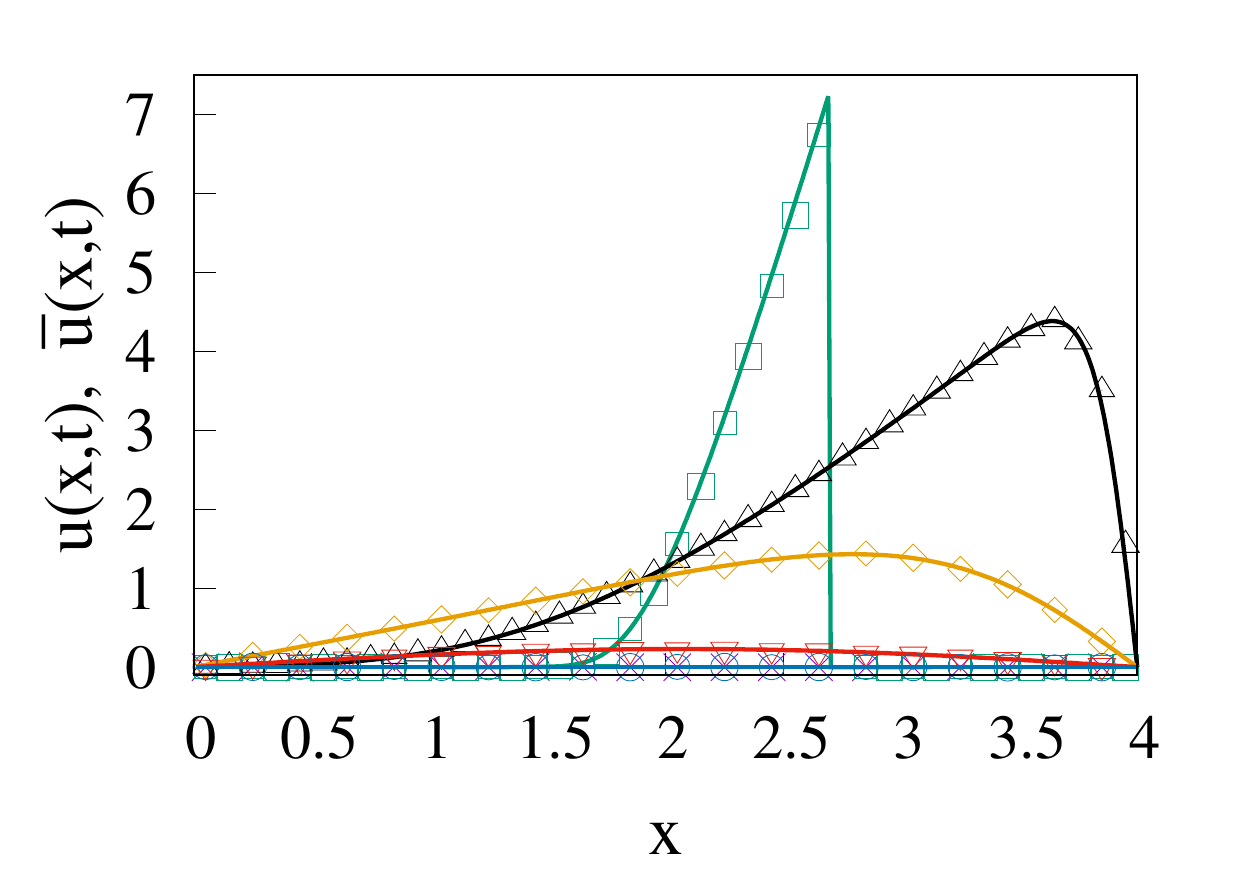}
\includegraphics[width=5.8cm,angle=0]{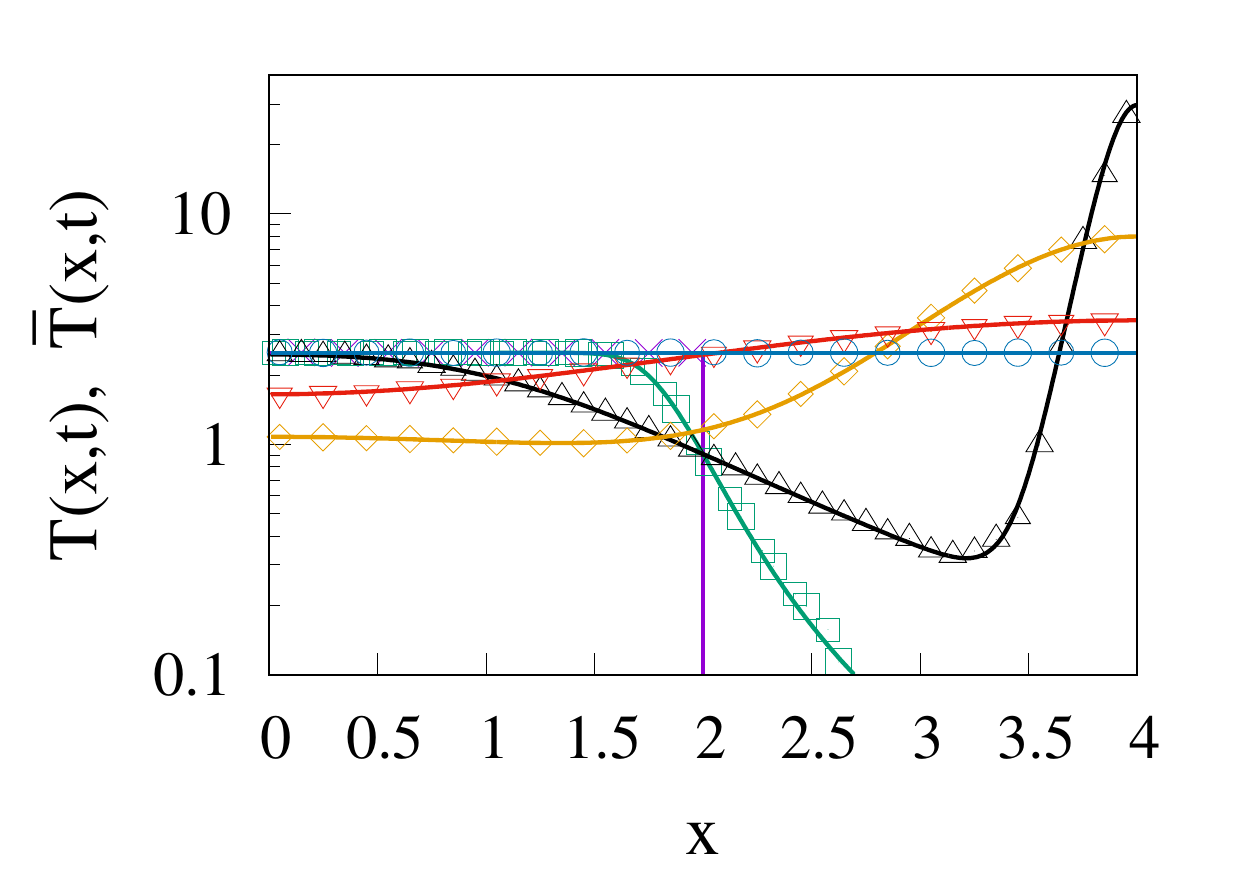}
\caption{Plot of the spatial profiles of the three conserved fields density $\rho(x,t)$, velocity $u(x,t)$, and temperature $T(x,t)$  at different times $t=0$ (magenta crosses), 0.1 (green empty squares), 0.4 (black empty triangles), 1 (yellow empty diamonds), 2 (red inverted empty triangles), and 4 (blue empty circles) obtained from simulation of a single typical microstate of $N=10^7$ particles. The density is normalized by the mean value $\rho_0=N/L$. Initial configuration is one realization of the canonical ensemble for particles in the left half $(0,L/2)$ with $L=4$. The initial positions of the particles are distributed uniformly between $(0,L/2)$ and the initial velocities are drawn from Maxwell distribution given by Eq.~\eqref{Maxwell} with  canonical temperature $T_0=2.5$.  The solid lines are analytically obtained fields $\bar{\rho},\bar{u}=\bar{p}/\bar{\rho},$ and $\bar{T}$ given by Eqs.~(\ref{eq_density_field}), (\ref{eq_momentum_field}), (\ref{eq_energy_field}) (see App.~\ref{sec:appendix-field} for details). The excellent agreement between the empirical densities and the mean densities once again establish typicality.}
\label{fig3}
\end{center}
\end{figure*}
\begin{figure}
\begin{center}
\leavevmode
\includegraphics[width=8.5cm,angle=0]{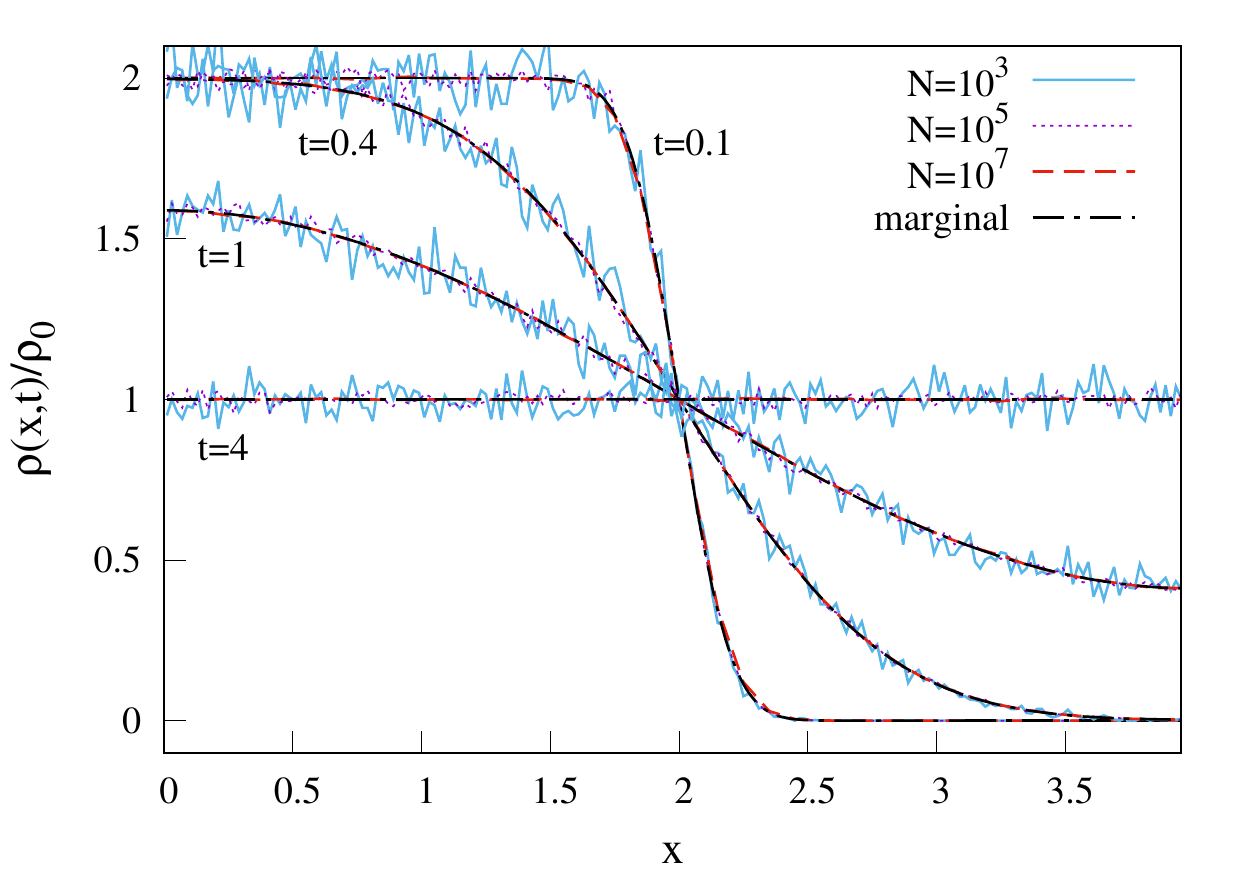}
\caption{Testing the dependence of typicality on the number of particles $N$. We plot the empirical density $\rho(x,t)$ (normalized by $\rho_0=N/L$) for different $N$ and at different times, along with the mean density. The agreement of the empirical profiles with the averaged ones (dashed-dotted lines) becomes better as $N$ is increased.}
\label{typical-density-diffN}
\end{center}
\end{figure}

\begin{figure}
\begin{center}
\leavevmode
\includegraphics[width=8.5cm,angle=0]{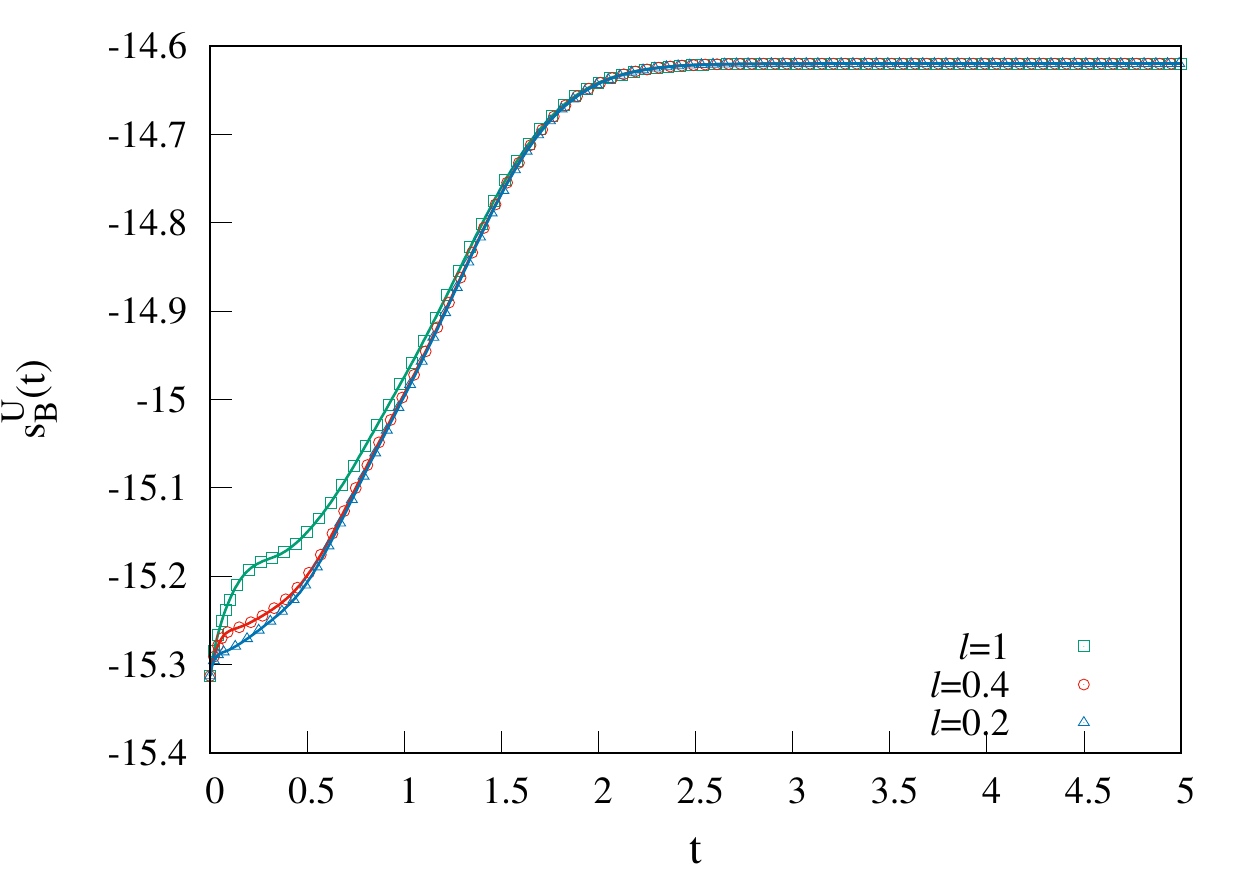}
\caption{Plot of entropy per particle, $s^U_B$, as a function of time during free expansion. Parameter values are: $L=4, N=10^5, T_0=2.5$. We consider the same single initial configuration as in Fig.~\eqref{fig1}. The different plots correspond to partitions of size $\ell=0.2,0.4,1$. Unlike in Fig.~\eqref{fig2}, here we see a monotonic increase and a convergence of the growth rate on decreasing $\ell$. The solid lines correspond to the mean field analytic profiles and we find very good agreement with the entropy computed from the empirical fields.}
\label{fig4}
\end{center}
\end{figure}

\subsection{Choice II of the macrovariables}

\subsubsection{Numerical results}
 We again start from a typical single realization  with $N=10^7,~L=4$ and $T_0=2.5$ (the same as that used in Fig.~\eqref{fig1}). In this case we partition the box into $K=40$ cells each of size $\ell=L/40=0.1$ and calculate the corresponding  empirical density $\rho(x,t)$, velocity $u(x,t)=p(x,t)/\rho(x,t)$ and energy $e(x,t)$ fields. Suppressing the time dependence, we have that $\rho_a = \rho(x_a), p_a = p(x_a)$, and $e_a = e(x_a)$, $x_a \in \delta_a$, with corresponding temperature field $T(x) = 2e(x)/\rho(x) - u^2(x)$.   In Fig.~\eqref{fig3}, we plot these fields at different times. The solid lines are the analytically obtained averaged fields $\bar{\rho},~\bar{u},$ and $\bar{T}$ given by Eqs.~(\ref{eq_density_field}), (\ref{eq_momentum_field}), (\ref{eq_energy_field}).  The details of the analytical calculation of mean fields  are provided in App.~\ref{sec:appendix-field}.  We find excellent agreement between the empirical and mean densities, as expected. We also find that at long times  these fields  converge to their equilibrium values given by the uniform fields $\rho(x)=\rho_0$,  $u(x)=0$ and $T(x)=T_0$. Unlike for the case of the $f$-macrovariable, here we do not see an oscillatory approach to the equilibrium state. In fact from the analytic results (see App.~\ref{sec:appendix-field}) one can see that the approach to equilibrium  at long times takes the form $A(x,t) - A_{\rm eq}(x) \sim B(x)~e^{-a t^2}$ with $a=T_0 \pi^2/(2L^2)$, where $A(x,t)$ can be any of the three fields  ${\rho},u,{T}$ discussed above, $A_{\rm eq}(x)$ represents its equilibrium value and $B(x)$ is some real known function.  
 Next,  we compute the  empirical density field for different values of $N$. In Fig.~\eqref{typical-density-diffN} we plot the evolution of $\rho(x,t)$ for the different values of $N$ and  compare them with the respective mean profiles $\bar{\rho}(x,t)$ at different times (black dot-dashed lines). We notice that the empirical density shows fluctuations for small $N$ which decrease for increasing $N$, leading to better agreement of the empirical profiles with the averaged ones.

We next insert these three fields into Eqs.~(\ref{Boltzmann_entropy2}) and (\ref{Boltzmann_entropy3}) to obtain the intensive empirical entropy  $s^U_B(t)$.
 In Fig.~\eqref{fig4} we plot  $s^U_B(t)$ with time $t$ for different cell sizes $\ell$. The solid lines correspond to theoretical computation of $s^U_B(t)$ using  the analytical expressions of the mean fields $\bar \rho(x,t),~\bar p(x,t)$ and $\bar e(x,t)$ given in App.~\ref{sec:appendix-field}. In this case we see that the increase of $s^U_B(t)$ is  monotonic and the entropy growth rate converges as we decrease the cell size $\ell$. The final increase of entropy is again equal to $\ln (2)$, as expected.

\subsubsection{Entropy increase for $S^U_B$ and hydrodynamics} 
\label{sec:discussion}
 We now explore the connection between the increase of the entropy $S^U_B$  and the behavior of the $U$-macrovariables in the hydrodynamic limit. It is believed that the Euler equations for the three conserved fields describe, in a suitable regime,   the hydrodynamics of a one-dimensional fluid of interacting particles.
At the level of the Euler equations there is no entropy increase. While this is well known, we provide an argument for it here, since we will need to refer to the argument later. So consider the  one-dimensional Euler equations: 
\begin{subequations}
\label{EulerAll}
\begin{align}
\label{Euler_R_1}
&\partial_t \rho+\partial_x(\rho u)=0, \\
\label{Euler_P_1}
&\partial_t (\rho u) + \partial_x(\rho u^2+P)=0, \\
\label{Euler_E_1}
&\partial_t \left(\rho \tilde{e} + \frac{1}{2} \rho u^2 \right) + \partial_x \left[ u \left( \rho \tilde{e}+ \frac{1}{2} \rho u^2 +P \right) \right]=0,
\end{align}
\end{subequations}
where $\tilde{e}(x,t)=e/\rho-u^2/2$  is the internal energy per particle and the pressure, for an ideal gas system is given by $P=\rho T$. These equations can be written in the form
\begin{subequations}
\begin{align}
\label{Euler_R_2}
&\frac{D\rho}{Dt}+ \rho \partial_x u = 0,\\
\label{Euler_P_3}
&\frac{Du}{Dt} + \frac{1}{\rho} \partial_xP =0, \\
\label{Euler_E_2}
&\frac{D\tilde{e}}{Dt} +\frac{1}{\rho} P{\partial_x u} =0,
\end{align}
\end{subequations}
where  $D/Dt=\partial_t + {u \partial_x}$ denotes the advective derivative.
\begin{figure}
\begin{center}
\leavevmode
\includegraphics[width=8.5cm,angle=0]{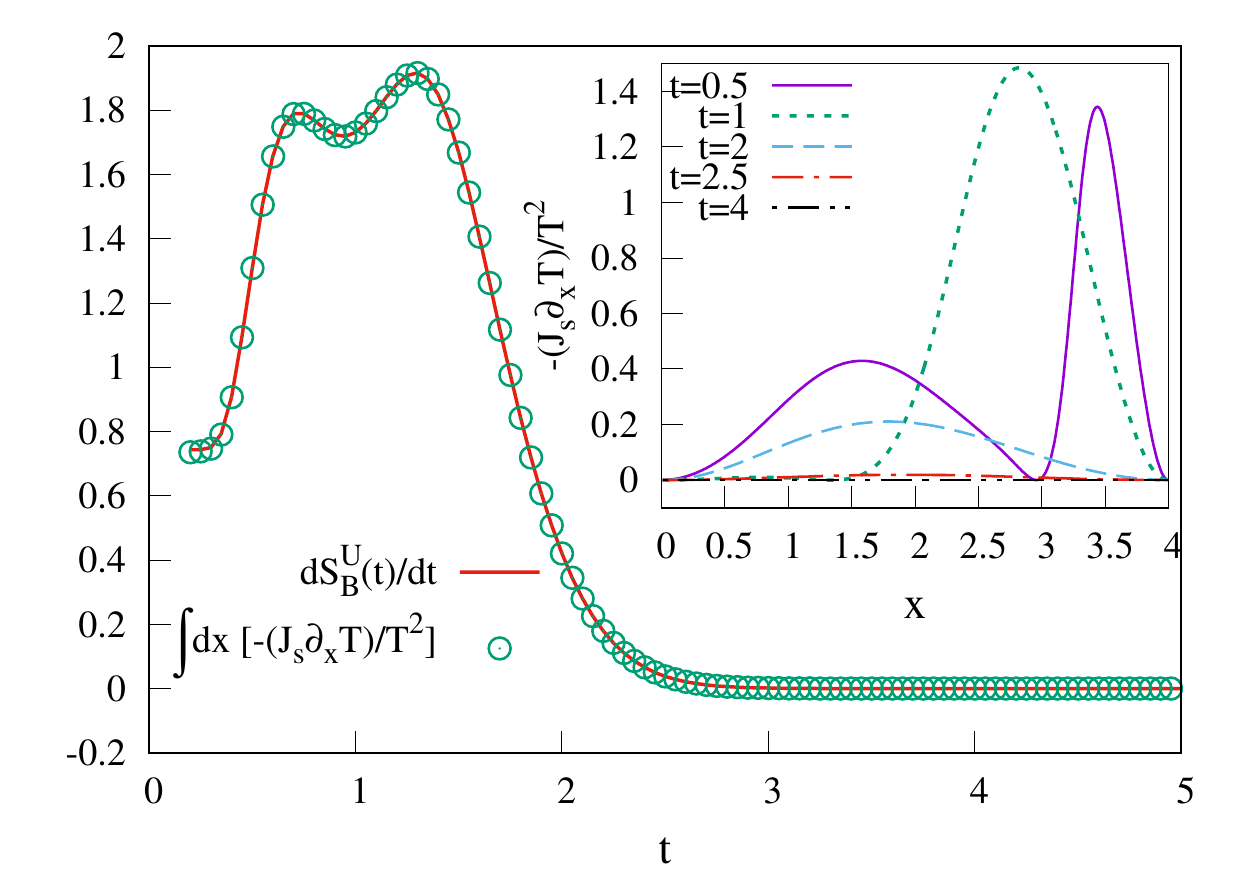}
\caption{Verification of entropy production rate as in Fig.~\eqref{fig4} using Eq.~\eqref{entropy_production}. The red line is the LHS of Eq.~\eqref{entropy_production} where $S^U_B(t)$ is calculated from the definition given by Eq.~\eqref{Boltzmann_entropy2} and the green points are the RHS of Eq.~\eqref{entropy_production} calculated from hydrodynamics. Inset: Plot of the integrand in Eq.~\eqref{entropy_production} with space $x$ at different times.}
\label{entropy_compare}
\end{center}
\end{figure}
Now we use the Euler hydrodynamic equations along with Clausius' laws of thermodynamics to determine the entropy production rate in the slowly evolving local equilibrium state. Clausius' laws of thermodynamics provide a well-known thermodynamic relation for the ideal gas, given by $TdS=d\tilde{E}+PdL$, where $S$ is the Clausius entropy, $\tilde{E}$ is total internal energy and $L$ is the volume. Applying this relation to a small volume $\ell$ with a fixed number of particles $n_\ell$  we find, after some manipulations, 
\bea \nonumber
&& Tds=d\tilde{e}+Pd(\ell/n_\ell), \\ \nonumber
&&{\rm hence,}~~ \frac{Ds}{Dt}=\frac{1}{T} \left[ \frac{D\tilde{e}}{Dt}-\frac{P}{\rho^2}\frac{D\rho}{Dt} \right],
\eea
with $s(x,t)$ being the entropy per particle. From Eqs.~(\ref{Euler_R_2}) and (\ref{Euler_E_2}) we then immediately obtain that $\frac{Ds}{Dt}=0$. The total entropy $S(t)=\int_0^L \rho s(x,t)~ dx$ also remains constant, since
$dS/dt = -\int_0^L dx ~ \partial_x ({\rho us}) = 0$,  using the boundary conditions ${u(0,t)=u(L,t)=0}.$ The standard mechanisms of entropy  growth in the hydrodynamic description are either additional  dissipative (Navier-Stokes-Fourier) terms or  the formation of shocks. 

\begin{figure*}
\begin{center}
\leavevmode
\includegraphics[width=15.5cm,angle=0]{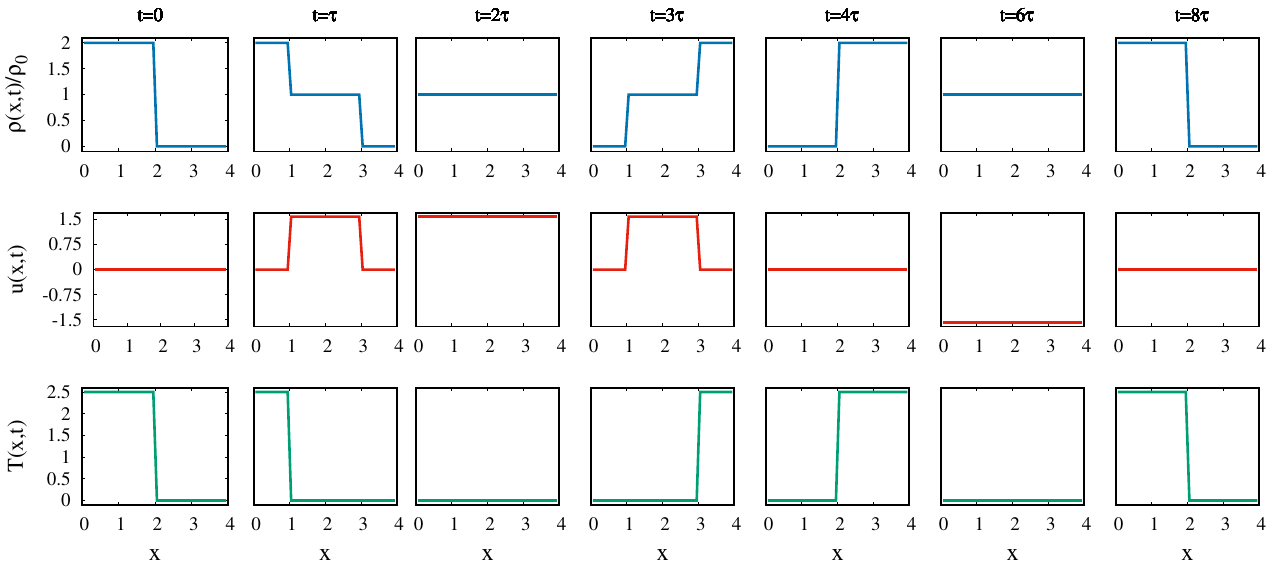}
\caption{Plot of the  spatial profiles of the three conserved fields density $\rho(x,t)$, velocity ${u(x,t)}$, and temperature $T(x,t)$  at different times obtained from simulation of a single ``atypical'' configuration with a binary choice of velocities. The initial positions of the particles ($N=10^7$) are distributed uniformly between $(0,L/2)$ with $L=4$, the initial velocities of odd particles are set $v_0$ and of even particles are set to $-v_0$. For comparison with previous cases we  chose $v_0=\sqrt{T_0}$ with $T_0=2.5$. The profiles repeat themselves after a time period $8\tau$ with $\tau=L/(4v_0)$ and thus the system does not reach an equilibrium at large time in this `atypical' case. We have used grid size $\ell=0.1$.}
\label{fig5}
\end{center}
\end{figure*}

\begin{figure}
\begin{center}
\leavevmode
\includegraphics[width=8.5cm,angle=0]{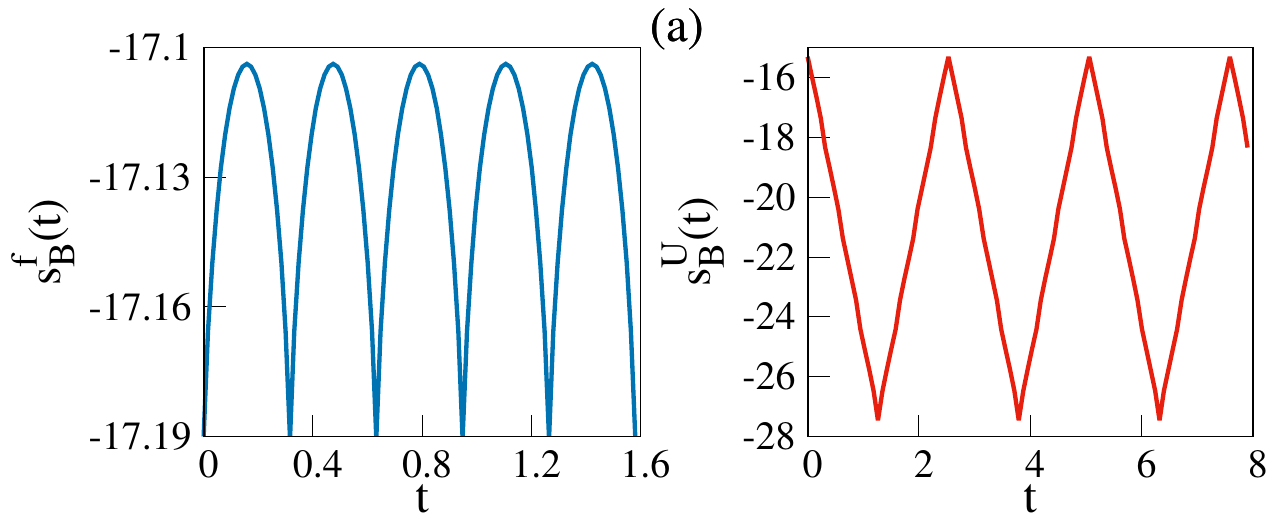}
\includegraphics[width=8.5cm,angle=0]{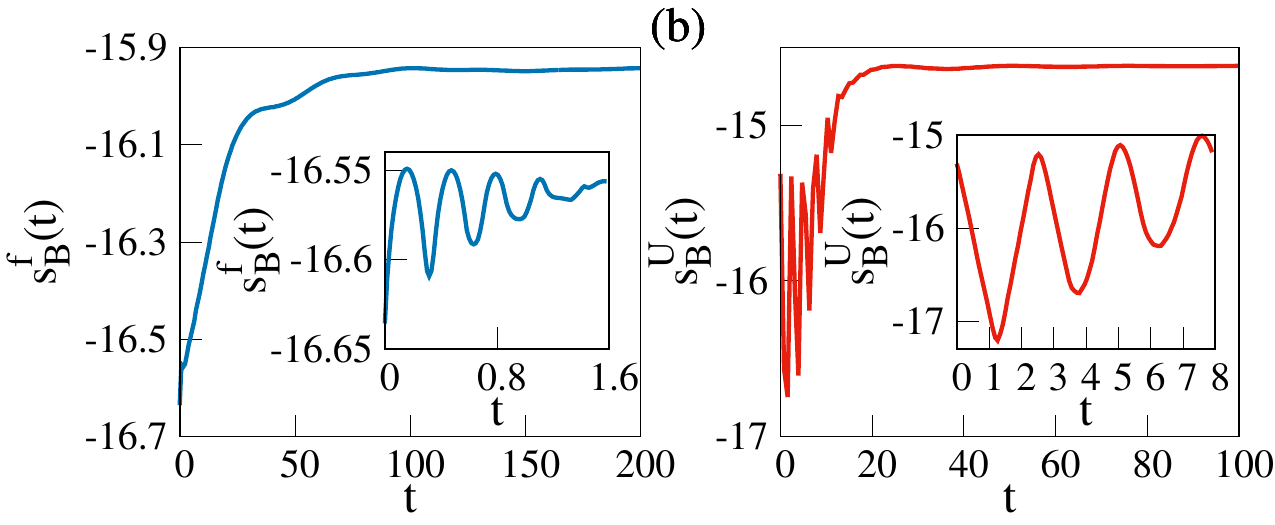}
\caption{Panel (a): Plot of $s_B^f(t)$ and $s_B^{U}(t)$ for the `atypical' initial configuration with binary velocities, and other parameters considered in Fig.~\eqref{fig5}. The periodic oscillation in both cases imply that the system never reaches equilibrium. For the left figure we have used grid size $\Delta x=\Delta v=0.5$ and for the right figure we have used $\ell =0.1$. Panel (b): In this case we consider a small perturbation of the binary velocity initial conditon in (a).  We choose  $v_i=v_0+ r_i$ for odd particles and  $v_i=- v_0+ r_i$ for even particles, where $r_i$ are iid random variables chosen from the uniform distribution over  $(-\epsilon,\epsilon)$ with $\epsilon = 0.1 v_0$.  We find that,  unlike the recurrent behaviour observed in (a), both the entropies increase and saturate to a final value with a final change of $\ln 2$. The insets in (b) show the early time evolution, where we see decaying oscillations, reminiscent of the recurrent ones seen in (a).}
\label{fig5a}
\end{center}
\end{figure}

We now discuss entropy production in our non-interacting gas using a similar description, keeping in mind that we now do not expect a closed set of hydrodynamic equations with the three fields. In fact  from Eq.~\eqref{feq} we see that the conserved densities $\rho(x,t)=N \int_{-\infty}^\infty dv \tilde{f}(x,v,t)$, $\rho(x,t) u(x,t) =N \int_{-\infty}^\infty dv ~v \tilde{f}(x,v,t) $ and $e(x,t)=N \int_{-\infty}^\infty dv (v^2/2) \tilde{f}(x,v,t)$ satisfy continuity equations. The first two of these are precisely the  Euler equations in  Eq.~\eqref{Euler_R_1} and \eqref{Euler_P_1}  whereas  the third equation for the energy field $e(x,t)$ is different from Eq.~\eqref{Euler_E_1}~\cite{frisch1958approach}. Formally the energy field satisfies the exact conservation equation
\bea
&& \partial_t  e+ \partial_x J=0, \\
&& {\rm where}~~J(x,t)= \frac{1}{2}\int_{-\infty}^\infty dv  ~v^3 N\tilde{f}(x,v,t)
\eea
is the energy current density. It is instructive to rewrite this equation in the following form: 
\begin{align}
\label{Euler_E_new}
&\partial_t e + \partial_x[{u (e + P)}]=-\partial_x J_s, \\
&{\rm where}~J_s(x,t)= J- {u  (e + P) }\label{eq_W}
\end{align}
is the current after subtracting the reversible Euler part. This current, $J_s$, can be interpreted as a ``heat'' current. Then, repeating the steps as before we find that the entropy production rate is finite and given by
\bea
\label{entropy_production}
\frac{d S(t)}{dt} = -\int_0^L dx ~ \frac{\partial_x J_s}{T}=-\int_0^L dx \frac{J_s \partial_x T}{T^2},
\eea
where in the last step we used the fact that the current vanishes at the boundaries. 
For our system we can compute the fields $J_s(x,t)$ and $T(x,t)$ directly from the exact solution of the microscopic dynamics and thereby compute the entropy production rate from the above equation.  {In Fig.~(\ref{entropy_compare}) we compare this  with the entropy production rate obtained from the definition given by Eq.~\eqref{Boltzmann_entropy2} and find perfect agreement between the two. In addition, as shown in the inset of Fig.~(\ref{entropy_compare}), we find that the integrand is non-negative everywhere (although we are not able to prove it explicitly) which leads to a non-negative entropy growth after integration. Note that for generic interacting non-integrable systems, the term $J_s$ should be expressible in terms of the three basic fields and in fact given by the Fourier's law $J_s = -\kappa \partial_x T$. This form would then guarantee non-negativity of the entropy production rate.}

We briefly comment on the growth of $S_B^f$. For the case of our non-interacting gas, Eq.~\eqref{feq} or Eq.~\eqref{Feq} are analogous to the Euler equations and  the growth of $S_B^f$ was  purely a result  of the discretization of $\mu$-space. Another well known trivially integrable system is the harmonic chain. The Euler equations
for this system were written in \cite{dobrushin1986} where it was  also noted that a finite space-time scaling parameter led to a Navier-Stokes type correction term~\cite{dobrushin1988}. Interestingly, for the disordered harmonic chain,  closed form Euler equations can be written for just the stretch and momentum variables, even though the system has a macroscopic number of conserved quantities~\cite{bernardin2019}.

\begin{figure*}
\begin{center}
\leavevmode
\includegraphics[width=5.8cm,angle=0]{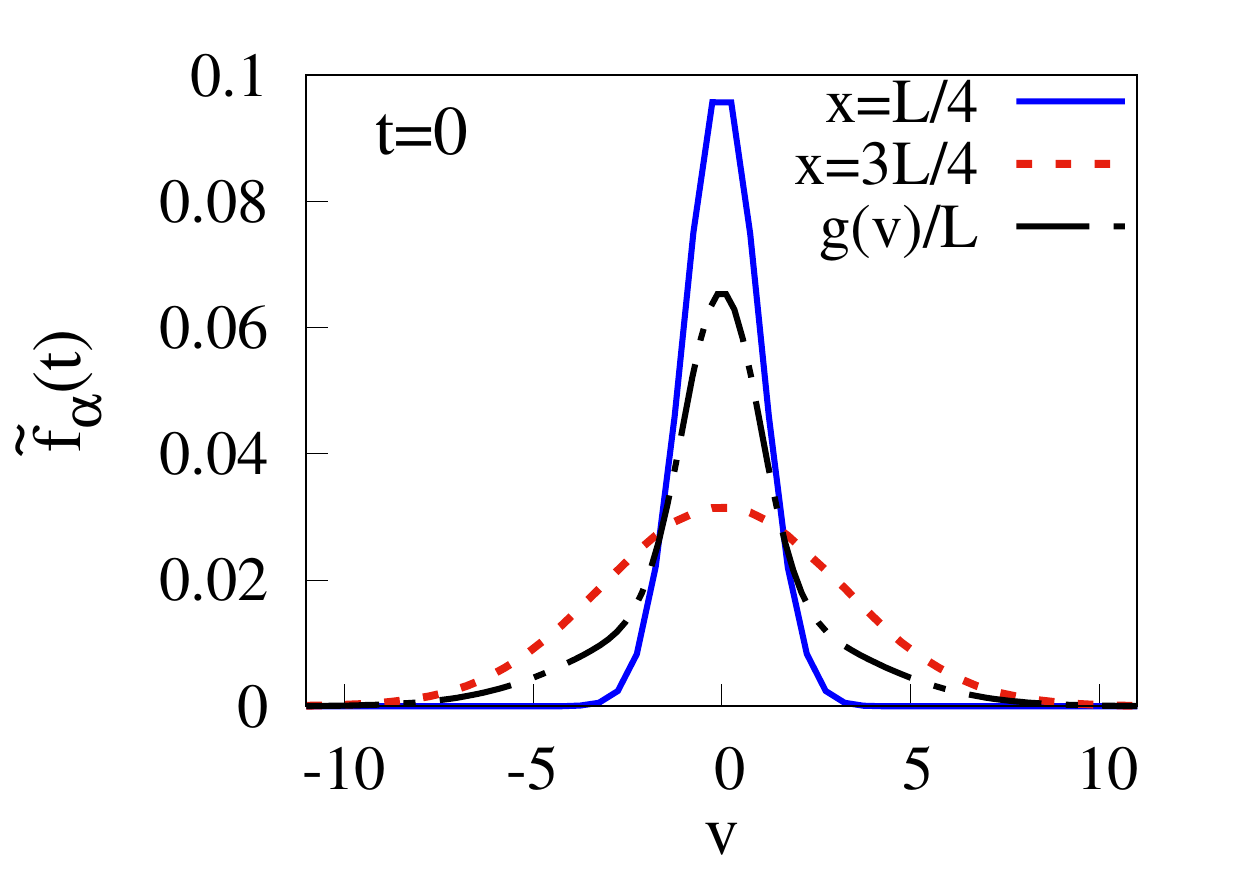}
\includegraphics[width=5.8cm,angle=0]{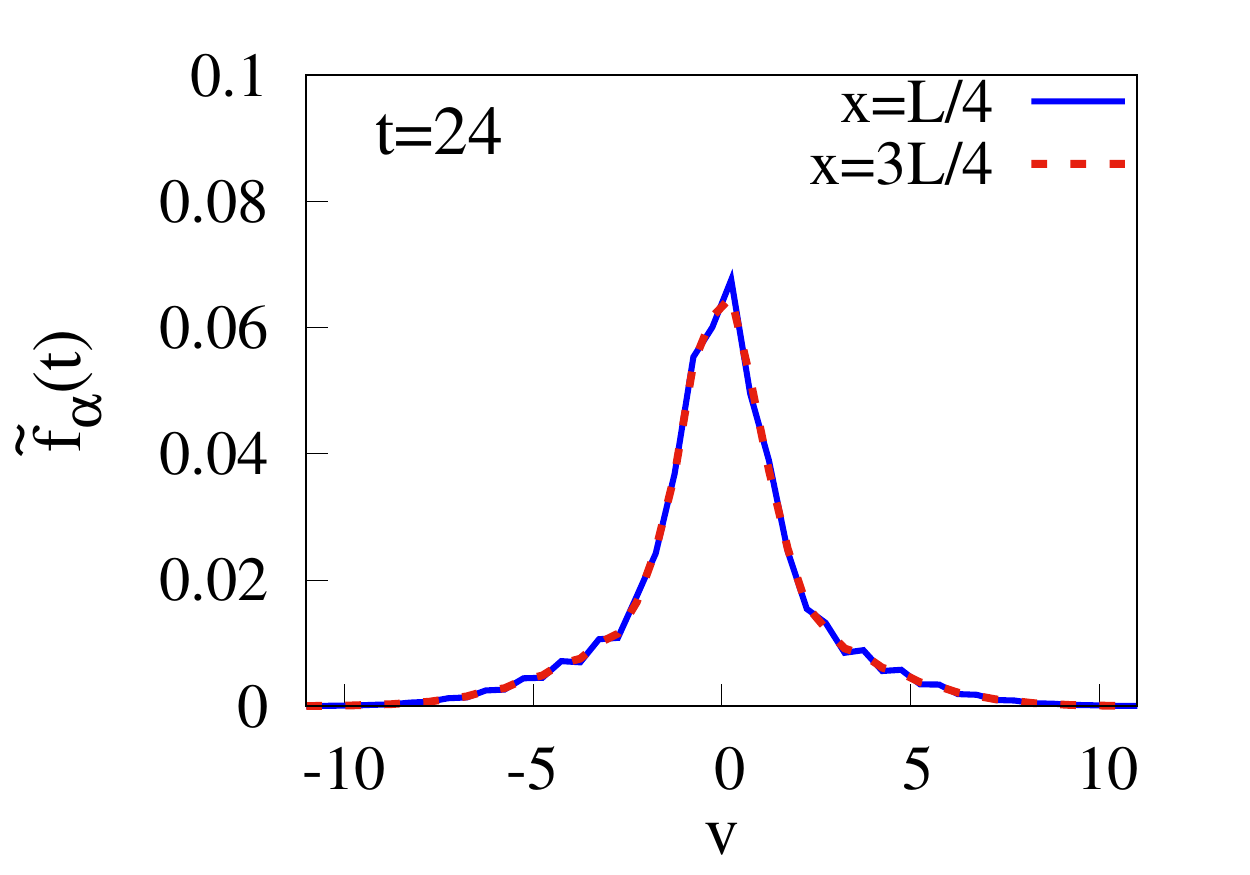}
\includegraphics[width=5.8cm,angle=0]{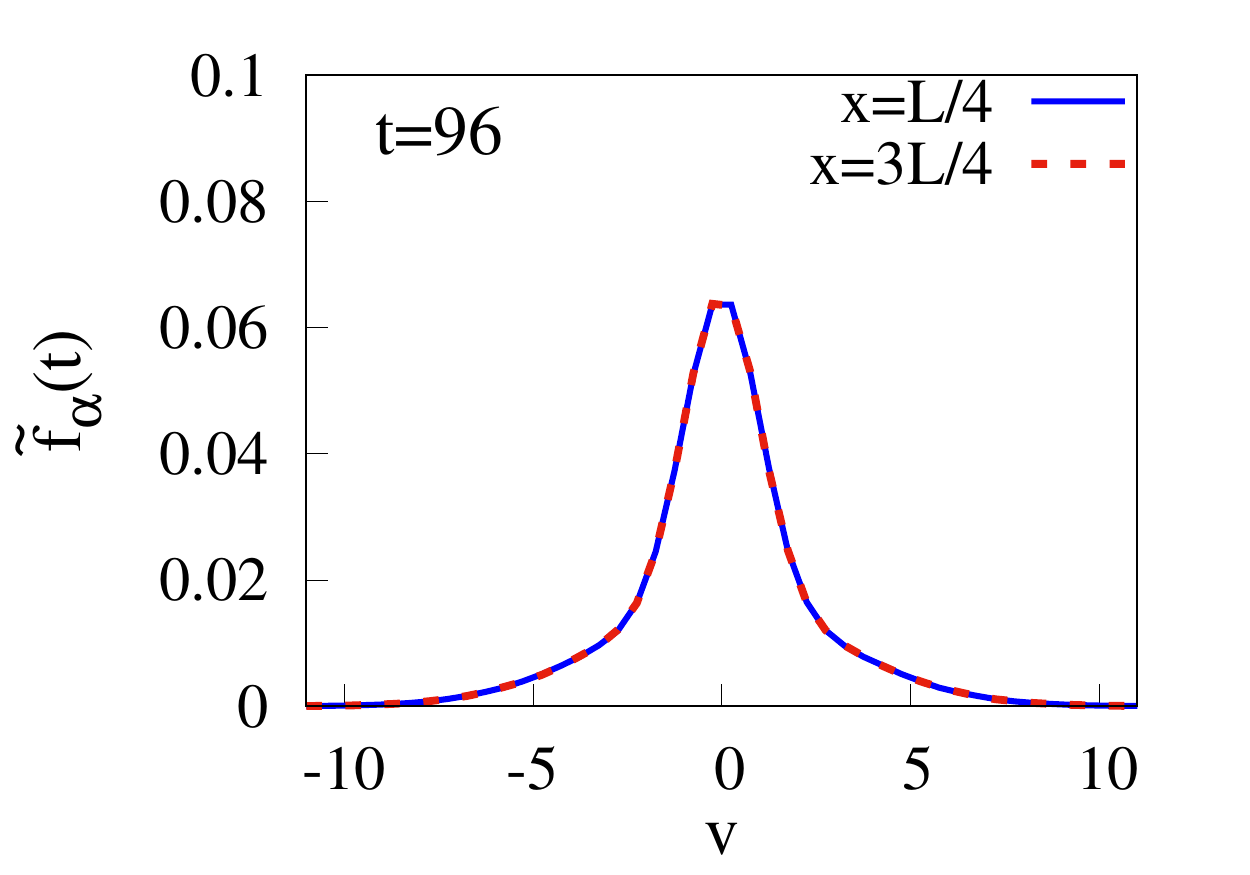}
\caption{Plot of evolution of the empirical  particle density $\tilde{f}_\alpha(x,v,t)=f_\alpha(x,v,t)/N$, starting from a single two-temperature initial microstate, at two spatial locations $x=L/4$ and $x=3L/4$. The initial position of $N=10^7$ particles are distributed uniformly within $(0,L)$ with $L=4$, while the initial velocities of the particles in the left and right halves are drawn from Maxwell distributions at temperatures $T_{L}=1$ and $T_{R}=10$, respectively. The  grid size was taken as $\Delta x=\Delta v=0.5$. At $t=0$ the empirical density $\tilde{f}_\alpha(x,v,t)$ is Maxwellian with $T_0=1$ and $T_0=10$ at $x=L/4$ and $x=3L/4$, respectively. As time evolves, the empirical density at any position gets contribution from particles originating initially from both the Maxwell distributions. After a large time, the distribution $\tilde{f}_\alpha(x,v,t)$ is seen to approach the form $g(v)/L$ (shown by the black dash-dot line in the left-most panel), where $g(v) = [{g_{\rm eq}(v,1)+g_{\rm eq}(v,10)}]/{2}$ and $g_{\rm eq}(v,T_0)$ is given in Eq.~\eqref{Maxwell}.}
\label{two-temp-f}
\end{center}
\end{figure*}

\begin{figure*}
\begin{center}
\leavevmode
\includegraphics[width=5.8cm,angle=0]{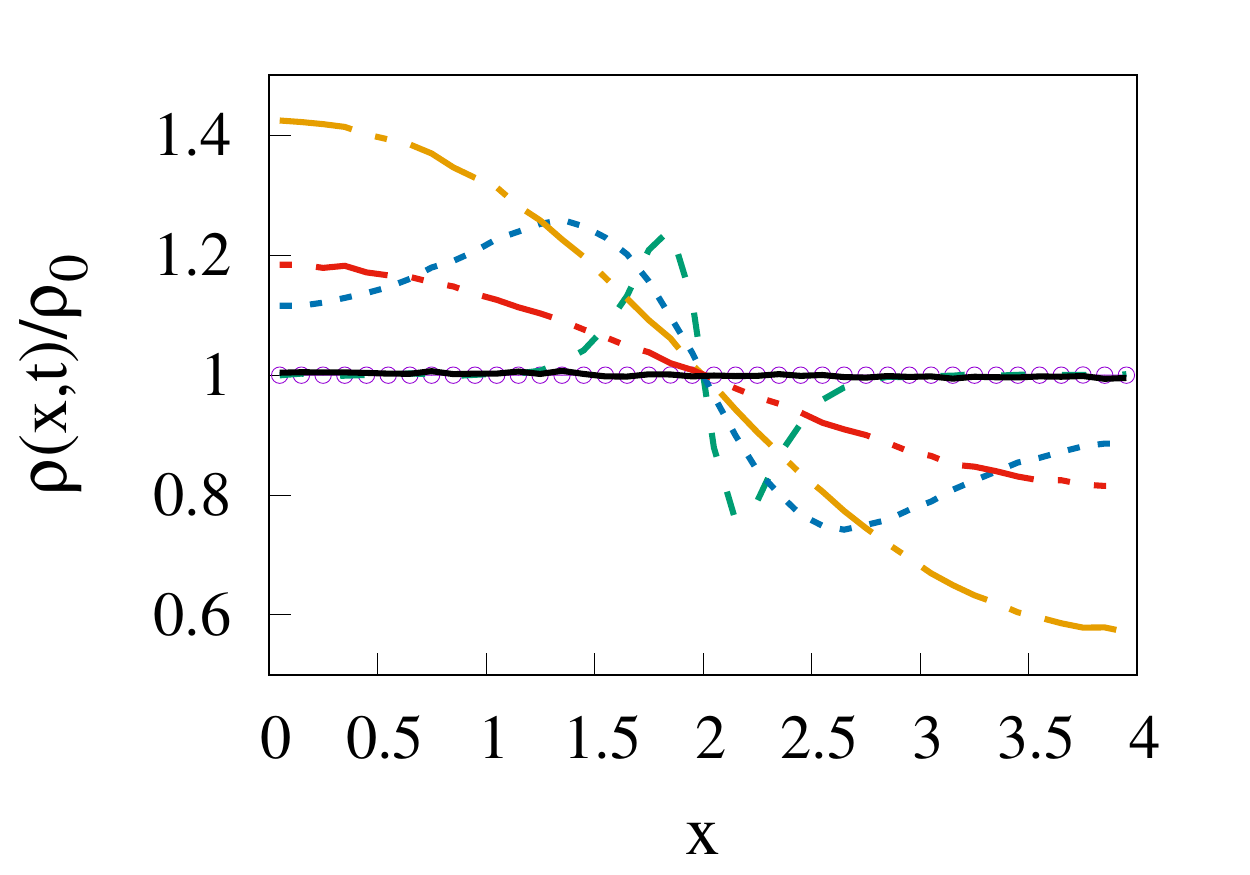}
\includegraphics[width=5.8cm,angle=0]{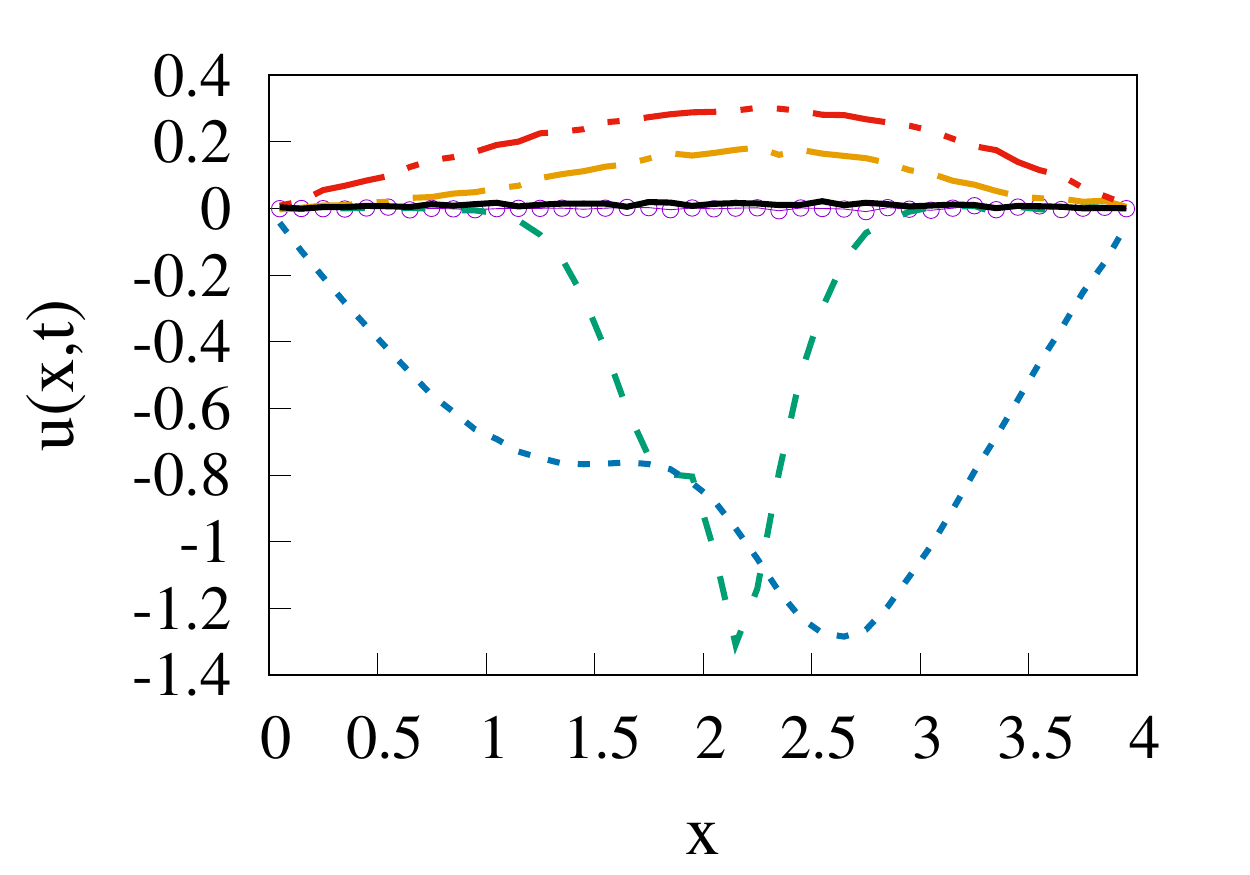}
\includegraphics[width=5.8cm,angle=0]{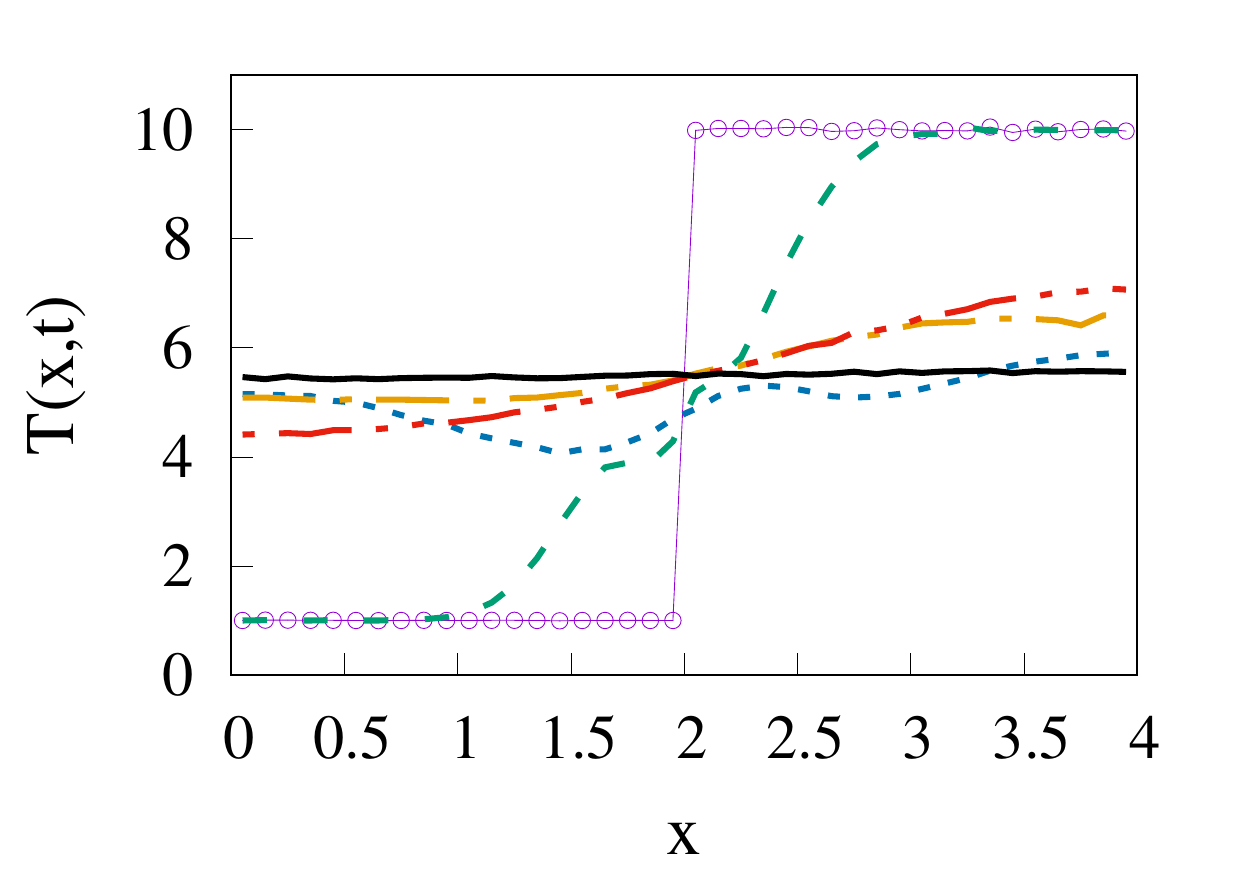}
\caption{Plot of the spatial profiles of the three conserved fields density $\rho(x,t)$, velocity ${u(x,t)}$, and temperature $T(x,t)$ for the `two-temperature case' at different times $t=0$ (magenta circles), 0.1 (green dashed lines), 0.4 (blue dotted lines), 1 (yellow dash-dot lines), 2 (red dash-dot-dot lines), and 4 (black solid lines) obtained from simulation of a single initial microscopic configuration with $N=10^7$ particles. The initial condition is the same as that used in Fig.~\eqref{two-temp-f}. Notice that after a highly nontrivial evolution, all three profiles become flat and thus the system reaches the equilibrium state at long times.}
\label{two-temp-profile}
\end{center}
\end{figure*}

\begin{figure}
\begin{center}
\leavevmode
\includegraphics[width=8.5cm,angle=0]{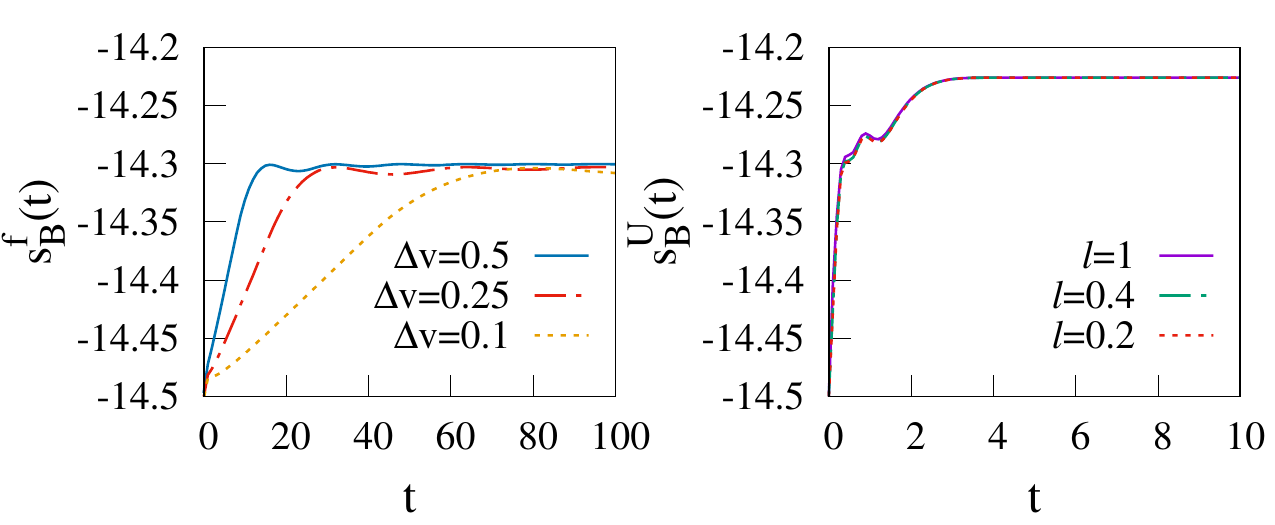}
\caption{Plot of $s_B^f(t)$ and $s_B^{U}(t)$ obtained from simulation of a single realization starting from the two-temperature initial configuration. The initial condition is the same as that used in Fig.~\eqref{two-temp-f}. We have used several grid sizes  for computing $s^f_B$ and $s^U_B$. While both entropies saturate at large times, only the final value of $s^U_B$ corresponds to the thermodynamic equilibrium value. Since the final velocity distribution at long times is not Maxwellian, the saturation value of $s^f_B$ does not correspond to the equilibrium value. Observe that the growth is non-monotonic in both cases. We also see convergence of $s^U_B$ with coarse-graining scale $\ell$ while for $s^f_B$ we again see a decreasing growth rate with decreasing $\Delta v$.}
\label{two-temp-sf}
\end{center}
\end{figure}

\section{Other initial conditions}
\label{sec:atypical}

So far we have considered a single typical initial condition for a macrostate in which  the initial positions are uniformly distributed in $(0,L/2)$ and the initial velocities are chosen from a (uniform) Maxwell distribution. We found that at large times the system goes to equilibrium, with the profiles of the conserved fields becoming flat and the corresponding entropy $s^U_B(t)$ reaching a steady value. It is also interesting to study the evolution for a single  initial condition, atypical for all the particles being on the left side. 

To do that we first consider a single configuration of $N=10^7$ particles initially in the left half $(0,L/2)$ distributed uniformly. The  initial velocities of odd particles are set to $v_0=\sqrt{T_0}$ and that of even particles are set to $v_0=-\sqrt{T_0}$, with $T_0=2.5$. Interestingly, in this case, each particle comes back to its original position with its original velocity periodically after a time period $2L/\sqrt{T_0}$. This recurrence is observed in Fig.~\eqref{fig5}, where we plot the profiles of the three conserved fields density $\rho(x,t)$, velocity $v(x,t)$, and temperature $T(x,t)$ at different times. We note that the profiles repeat themselves after a time period $8\tau$ with $\tau=L/(4\sqrt{T_0})$.  Thus, unlike for the typical initial configuration in Fig.~\eqref{fig3}, for this atypical initial condition the system never settles down into an equilibrium state for either of our two choices of macrovariables.
 We have also looked at the evolution of the entropies $s_B^f(t)$ and $s_B^{U}(t)$ for this atypical initial configuration in Fig.~(\ref{fig5a}a), where we find, of course, that the entropy in both cases keeps oscillating for all time.

We now demonstrate that even a small perturbation of this atypical initial condition changes the entropy evolution drastically --- from recurrent to irreversible growth (for times that are not extremely large). For this we add a small random perturbation to the previously discussed binary velocity setting. We choose  $v_i=v_0+ r_i$ for odd particles and 
 $v_i=- v_0+ r_i$ for even particles, where $r_i$ are iid random variables chosen from the uniform distributions over  $(-\epsilon,\epsilon)$ with $\epsilon \ll v_0$. The outcome is remarkable -- as seen in Fig.~(\ref{fig5a}b), both the entropies $s_B^f(t)$ and $s_B^{U}(t)$ now again increase by the amount $\ln(2)$ and finally reach equilibrium (though not their thermal equilibrium values). This clearly shows that the atypical initial microstate is  very special.  A slight perturbation makes it a more typical one  for which the entropy increases and we observe macroscopic irreversibility. In the insets of Fig.~(\ref{fig5a}b), we see oscillations of the entropy at early times, corresponding to some memory of the atypical initial condition.

The perturbed initial microstate just described is in fact still atypical, even for the $f$-macrostate to which it belongs---a consequence of the alternation of the sign of the velocity of successive particles. Nonetheless, the analysis of Sec.~\eqref{app:scaling} continues to  apply, providing an analytical demonstration of the irreversible behavior just described for the perturbed microstate.

We also consider another initial microstate corresponding to  an $f$-macrostate (choice-I) where the particles are uniformly distributed over the full box with zero momentum and with two different temperatures on the left and right half of the box, i.e., with the velocities chosen from the corresponding Maxwellians. In Fig.~\eqref{two-temp-f} we show the time-evolution of $f(x,v,t)$ while Fig.~\eqref{two-temp-profile} shows the evolution of the three fields -- density, velocity and temperature. In the former case we see that the long-time form of the single particle distribution is non-thermal, {i.e.} non-Maxwellian, as demonstrated and explained in Fig.~\eqref{two-temp-f}. On the other hand, the fields $\rho(x,t),~{u(x,t)}$ and $T(x,t)$ are converging to their expected thermal equilibrium values, as shown in Fig.~\eqref{two-temp-profile}.  In Fig.~\eqref{two-temp-sf} we compare the evolution of  $s^f_B$ and $s_B^U$ for the case where the left and right halves are initially at temperatures $T_{L}=1$ and $T_{R}=10$ respectively. We note that both $s^U_B$   and $s^f_B$ saturate at long times but the increase in entropy is less for $s^f_B$. This is because  the conserved fields evolve at long times to their thermodynamic equilibrium values, with uniform density, zero momentum and temperature $T=(T_{L}+T_R)/2$, with $s^U_B$ thus attaining the corresponding equilibrium value. On the other hand, the total velocity distribution does not evolve with time and hence remains non-Maxwellian  at all times. Thus $s^f_B$ saturates to a value lower than the equilibrium one. From Eq.~\eqref{fatn1} and the discussion in that section we find that the saturation value of $s^f_B$ is  given by  Eq.~\eqref{smax} 
with $\bar{g}(v)=[g_{\rm eq}(v,T_L)+g_{\rm eq}(v,T_R)]/2$. This agrees with the measured saturation value in Fig.~\eqref{two-temp-sf}. The invariant velocity distribution $\bar{g}(v)$ which is just the mean of two Maxwellians in fact  defines a corresponding generalized Gibbs ensemble (GGE) that describes the long-time equilibrium state of  the system. Note that the entropy growth for both definitions of entropy is non-monotonic, unlike what is seen for $s^U_B$ for free expansion.

\begin{figure}
\begin{center}
\leavevmode
\includegraphics[width=8.5cm,angle=0]{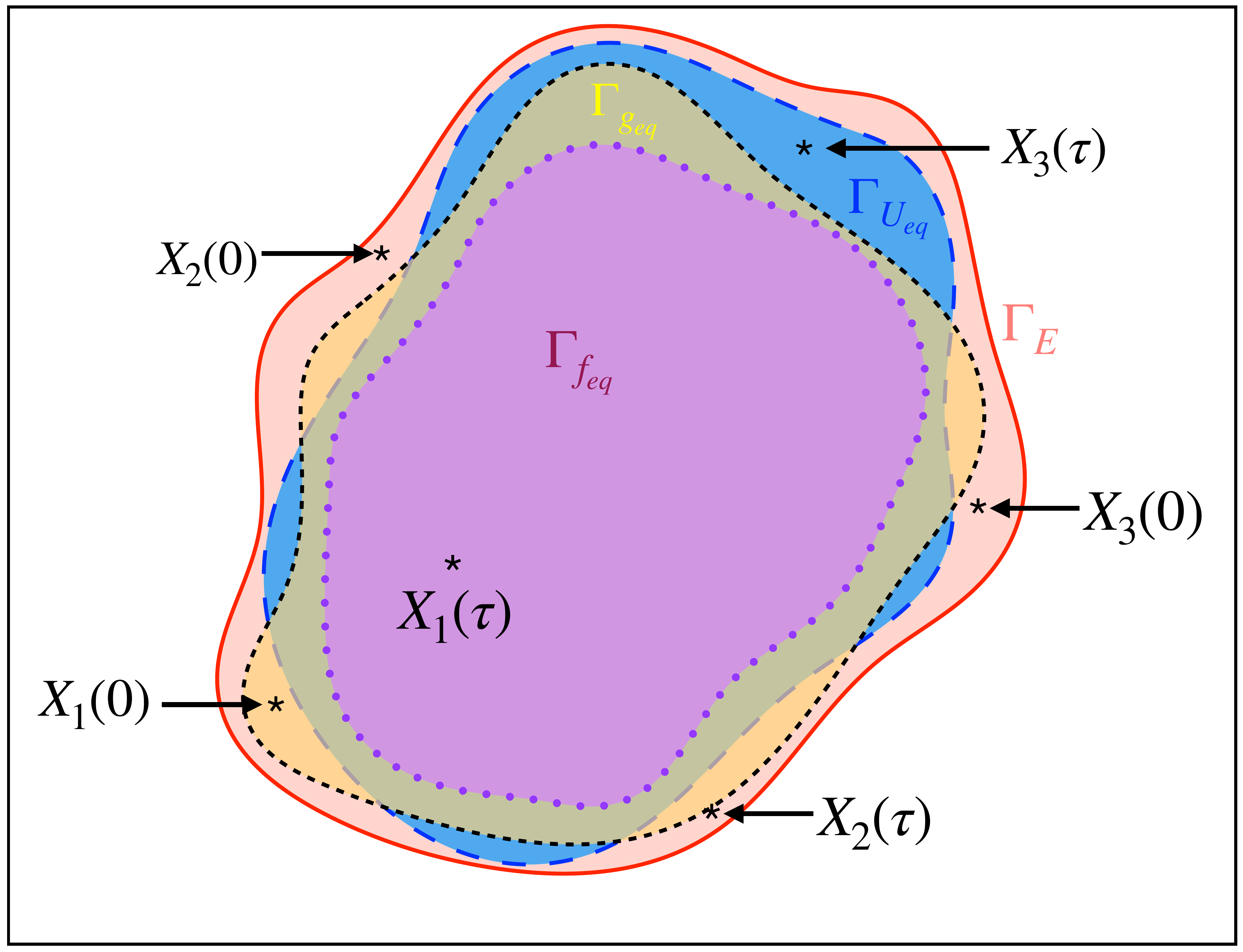}
\caption{A schematic showing the partition of the full phase space into subspaces defined by the equilibrium macrostates $f_{\rm eq}$ (boundary shown by dotted-line), $U_{\rm eq}$ (boundary with long-dashed line) and $g_{\rm eq}$ (boundary with dashed line). The points $X_1(0),X_2(0),X_3(0)$ correspond to the three initial microstates considered in our study. The point $X_1(0)$ corresponds to the  free expansion from thermal equilibrium, $X_2(0)$ corresponds to free expansion from the alternate velocity microstate and $X_3(0)$ corresponds to the uniform density non-Maxwellian (two-temperature) initial global velocity distribution. After a long time, $\tau$, these microstates,  $X_1(\tau),X_2(\tau),X_3(\tau)$, would remain in regions of phase space as shown in the figure ( making brief excursions out of these regions on Poincare recurrence time-scales).  We note that $X_1(\tau)$ will be contained in $\Gamma_{f_{\rm eq}}$, $X_2(\tau)$ is outside $\Gamma_{f_{\rm eq}} \cup \Gamma_{U_{\rm eq}} \cup \Gamma_{g_{\rm eq}}$, and $X_3(\tau)$ will end up in $\Gamma_{U_{\rm eq}}\setminus\Gamma_{f_{\rm eq}}$ (all points in $\Gamma_{U_{\rm eq}}$ that are not in $\Gamma_{f_{\rm eq}}$). }
\label{partition}
\end{center}
\end{figure}

\section{Geometrical overview}
\label{sec:geometric}
 Apart from the macrovariables, $U, f$, let us also define another one, corresponding to  the global velocity distribution: $g(v)=\int dx f(x,v)/N$. For the equal mass gas,  this  is a constant of the motion. Each of $U$, $f$ and $g$ define partitions of $\Gamma_E$ : $\Gamma_E=\{\Gamma_U\}=\{\Gamma_f\}=\{\Gamma_g\}$, where $\{...\}$ represents the collection of all possible macrostate values. The last, $\{\Gamma_g\}$, is a partition of  $\Gamma_E$  into sets invariant under the dynamics.  Each of the three partitions has a dominant set, $\Gamma_{U_{\rm eq}}$,  $\Gamma_{f_{\rm eq}}$,  and $\Gamma_{g_{\rm eq}}$, respectively. These are shown schematically in Fig.~\eqref{partition}. 
The macrostate $U_{\rm eq}$ corresponds to   uniform profiles of the conserved fields, $f_{\rm eq}$ corresponds to a macrostate with uniform density profile and Maxwellian velocity distribution, and $g_{\rm eq}$ corresponds to a global Maxwellian velocity distribution.
Note that the $f$ partition of $\Gamma_E$ is a refinement of the $U$ partition and also a refinement of the $g$ partition.

As shown in the figure, $\Gamma_{f_{\rm eq}}$ is the dominant set  in $\Gamma_{U_{\rm eq}}$ and in $\Gamma_{g_{\rm eq}}$ ,   while $\Gamma_{g_{\rm eq}}$ has tiny regions that are  outside $\Gamma_{U_{\rm eq}}$ and vice-versa. Any initial microstate, $X_1(0)$, inside $\Gamma_{g_{\rm eq}}$, such as the one chosen from thermal equilibrium in the left half of the box, will eventually be in the region $\Gamma_{f_{\rm eq}} \cap \Gamma_{U_{\rm eq}} \cap \Gamma_{g_{\rm eq}} =\Gamma_{f_{\rm eq}} $ which corresponds to ``complete'' thermal equilibrium. On the other hand typical microstates such as $X_3(0)$, chosen from outside of $\Gamma_{g_{\rm eq}}$ will end in $\Gamma_{U_{\rm eq}}$ but outside $\Gamma_{f_{\rm eq}}$ and so in this case we have restricted thermalization. This is  seen in Fig.~\eqref{two-temp-sf} where $s^U_B$ is seen to reach its equilibrium value while $s^f_B$ does not. Finally one has very special atypical microstates, $X_2(0)$, such as in the alternate velocity case considered in Fig.~\eqref{fig5a}, which remains outside $\Gamma_{f_{\rm eq}} \cup \Gamma_{U_{\rm eq}} \cup \Gamma_{g_{\rm eq}}$ and there is no thermalization at all. 
The above features are specific to our system, a non-interacting integrable model. For non-integrable models it is expected that almost any initial microstate would end in $\Gamma_{f_{\rm eq}}$ and the system would thermalize completely.

\section{Conclusion}
\label{sec:conclusions}

We summarize here our main findings. In this paper, we have studied  entropy increase during the free expansion of an ideal gas. In the microscopic description we start from an initial condition where the molecules are uniformly distributed in the left half of a box and the velocities are chosen from a thermal distribution. For this system we study the evolution of the Boltzmann entropy defined for a single microstate with two choices of macrovariables: the empirical single particle distribution $f_\alpha(t)$ [defined in Eq.~\eqref{falpha}] and the profiles of density, momentum and energy $U=[\rho(x),p(x),e(x)]$ (or equivalently $\rho,v,T$). The corresponding entropies are $s^f_B$ and $s^U_B$, respectively. In equilibrium, both these choices correspond to the thermodynamic entropy $S$. 
\begin{itemize}
\item  The time evolution of the empirical density $f(x,v,t)$ and the fields $\rho(x,t),{u(x,t)},T(x,t)$ were obtained for a  single realization, starting from a single typical initial microstate chosen from the equilibrium macrostate where all particles are in equilibrium in the left half of a box. We found that these agree, in the large $N$ limit, with the corresponding fields  $F_\alpha,\bar{\rho},{\bar{u}},\bar{T}$, obtained by \emph{averaging} over  initial microstates taken  from the relevant initial Gibbs distribution or, more or less equivalently, from the same initial macrostate.   For our model, the averaged fields can easily be  computed exactly. This demonstrates that  the evolution of the macrostates and the corresponding entropies, for single typical microstates (belonging to a macrostate $M_{\rm ini}$), agrees with   the evolution obtained after averaging over the ensemble of initial conditions (corresponding to the same initial macrostate $M_{\rm ini}$).
\item Both $s^f_B$ and $s^U_B$  increase with time and eventually reach the expected equilibrium value with an increase of $\ln(2)$ per particle.  However, while $s^U_B$ increases  monotonically with time,  $s^f_B$  shows oscillations which decay with time.

\item The  entropies are defined in terms of  coarse graining scales $| \Delta_\alpha |$ for $s^f_B$ and $\ell$ for $s^U_B$.  We find that the entropy production rate for  $s^U_B$ does not  depend much on decreasing grid-size $\ell$. On the other hand, $s^f_B$ decreases with decreasing  $\Delta v$. However, we find a remarkable scaling collapse of the data for $s^f_B(t)$ for different $\Delta v$ on plotting them as a function of the scaled time $\tau= t \Delta v/(2 L)$. We provide an analytic understanding of this collapse and compute the  scaling function. This also explains the  oscillations and in particular the observation that the system periodically goes to the maximum entropy state. 

\item The fact that the entropy production rate for $s^U_B$ does not depend  much on grid size in the ``continuum limit'' of  small grid size reflects a general feature of entropy production: it often is insensitive to the choice of macrovariables, which tend naturally to be chosen so as to be near the continuum limit of small grid size. This in part explains why the fact that the Boltzmann entropy depends upon, and is defined relative to, a choice of macrovariables and macrostates can often be ignored. However when the entropy production in the continuum limit vanishes, as it does in an ideal gas for the $f$-macrovariables, the dependence on grid size can no longer be ignored.

\item We showed that the entropy increase for $S^U_B$ can be related to a microscopic ``heat'' current, $J_s$, and formally the local entropy production rate can be written in the form $J_s \partial_x(1/T)$. This quantity is positive everywhere and its integral over all space gives a positive entropy production.

\item {\bf Other initial conditions}:  The results above are for the specific case of free expansion. We have also studied other initial conditions.  We considered one example (with an atypical initial microstate with alternate particles having different velocities, $\pm v_0$) where the system never reaches a steady state and the entropy shows persistent oscillations. Our second example involves particles initially distributed uniformly in the box but with a non-Maxwellian global velocity distribution. For this case we find that $s^U_B$ saturates to its equilibrium value at long times, while $s^f_B$ does not, corresponding to the observed fact that the macrovariables $f$ and $U$ evolve in this case to limiting values: $U$ evolves to its equilibrium value  $U_{\rm eq}$ while $f$ evolves to $g/L$, corresponding to the dominant $f$-macrostate given the total velocity distribution arising from $f_{\rm initial}$.
 \end{itemize}

Thus our study illustrates the crucial role of typicality, large numbers, and coarse-graining that lead to entropy increase (starting from a low entropy state),  irreversibility  and  approach to thermal equilibrium, even in a non-interacting integrable system. The ideas of ergodicity, interaction and chaos do not seem to be so relevant. 

That there is no change in the continuum limit $\mathcal{s}^f_B$, the entropy $s_B^f$ in the limit $\Delta x, \Delta v \to 0$, agrees computationally with what would be obtained from generalized hydrodynamics (GHD) of integrable systems \cite{Alvaredo_PRX2016,spohn2018,Nardis_PRL2018,Nardis_SPP2019,Nardis_JSM2022}.
According to GHD, the entropy is  defined as the integral of a  Gibbs entropy density for  local generalized canonical ensembles involving other conserved quantities in addition to the energy.   This is regarded as the fundamental entropy, and not, as we do,  as an idealization of the more fundamental entropy $s_B^f$ that depends on the choice of macrovariables and coarse graining.
Classical integrable systems contain a macroscopic number of conserved quantities and, according to GHD, the correct hydrodynamic description of an integrable system is given by a specification of all the conserved fields. One then finds that for non-interacting systems such as the ideal gas or a harmonic chain, the hydrodynamic equations are given by the (generalized) Euler equations,  which yield  no entropy production for the corresponding continuum entropy.

In fact, for the ideal gas, it is seen  that the empirical single particle density, $\tilde{f}(x,v)$,  has complete information about the conserved quantities and the hydrodynamic equations are equivalently given by Eq.~\eqref{feq}, implying constancy of the continuum entropy $\mathcal{s}_B^{f}(t)$ [see Eq.~\eqref{H-function3}]. On the other hand, if instead we consider the contiuum entropy for the $U$-macrostates, involving only three of the conserved quantities, we find entropy increase.  For ``interacting'' integrable models, such as the hard rod system and the Toda chain, it turns out that Navier-Stokes dissipative terms appear in the hydrodynamic equations, and hence there is continuum entropy generation even when the macrostate description includes all conserved quantities. We hasten to add, however, that unlike for GHD the choice of macrovariables required for a Boltzmann entropy need not be limited to those that are conserved. After all, the one-particle empirical distribution $f(x,v)$ involved in the Boltzmann equation for a hard sphere gas is certainly not conserved when collisions occur.

Thermalization in non-chaotic systems, such as the Toda chain, has been discussed in  other recent papers~\cite{baldovin2021,ganapa2020}. The  one-dimensional hard rod gas, an integrable model which however has a Navier-Stokes term in the hydrodynamic equations~\cite{Doyon_JSM2017},  would be another interesting case to study from the present perspective. It will also be interesting to extend the study of the present paper to  interacting systems such as the alternate mass hard particle gas  where the masses of alternate particles take different values. This non-integrable model has been studied extensively in the context of the breakdown of Fourier's law of heat conduction in one dimension \cite{Dhar_PRL2001,Grassberger_PRL2002,Casati_PRE2003,Cipriani_PRL2005} and verification of the hydrodynamic description~\cite{Spohn_JSP2014,Mendl_JSP2016,subhadip2021,ganapa2021}. For this case we expect important differences from the present integrable model even though the equilibrium thermodynamics of both are identical. For example we expect that the entropy production rate for $s_B^f$ at a fixed time should converge to a finite positive value even in the limit of vanishing grid-size, in contrast to what we see in Fig.~\eqref{fig2}.  

\section*{Acknowledgments}
 SC, AD and AK acknowledge support of the Department of Atomic Energy, Government of India, under Project No. RTI4001 and would also like to acknowledge the ICTS program on “Thermalization, Many body localization and Hydrodynamics (Code:ICTS/hydrodynamics2019/11)” for enabling crucial discussions related to this work.  SC, AD and AK thank Cedric Bernardin, Deepak Dhar, Francois Huveneers, Christian Maes, Stefano Olla, and Herbert Spohn for illuminating discussions. We thank Cedric Bernardin,  Vir Bulchandani, Benjamin Doyon, David Huse and Joel Moore for helpful comments on the manuscript. The work of JLL was supported by the AFOSR. The numerical simulations were performed on a Mario HPC at ICTS-TIFR.

\bibliography{HPG_entropy}

\appendix

\section{Exact solution for the evolution of macroscopic fields}
\label{sec:appendix-field}

Here we present exact results for the evolution of average fields corresponding to the empirical densities $f(x,p,t)$ and  $[\rho(x,t),p(x,t),e(x,t)]$. A similar study on a ring was done earlier in \cite{frisch1958approach,frisch1956}.

The mean density $\bar{\rho}(x,t)$, momentum $\bar{p}(x,t)$ and energy $\bar{e}(x,t)$ fields are defined as 
\bea \nn
\label{eq_d_field}
\bar{\rho}(x,t)= \left \langle \sum_{j=1}^N \delta(x_j - x) \right \rangle = \int dv~F(x,v,t), \\
\\ \nn
\label{eq_p_field}
\bar{p}(x,t)=  \left \langle \sum_{j=1}^N \delta(x_j - x) v_j \right \rangle = \int dv~v~F(x,v,t), \\
\\ \nn
\label{eq_e_field}
\bar{e}(x,t)=  \left \langle \sum_{j=1}^N \delta(x_j - x) \frac{v_j^2}{2} \right \rangle = \int dv~\frac{v^2}{2}~F(x,v,t). \\
\eea
Here $\langle ... \rangle$ denotes an average over  the  initial positions and velocities, $\{x_i(0),v_i(0)\}$, of all the particles, which are chosen from the canonical distribution at temperature $T$ with all particles in the left half $(0,L/2)$.   In the following, we study the evolution of these fields in time and space for this system with two reflecting boundaries at $x=0$ and $x=L$. 

 The canonical ensemble implies that we distribute the particles  uniformly in $(0,L/2$) and draw their velocities from the Maxwell's velocity distribution given by
\be
\label{Maxwell-a}
g_{\text {eq}}(v_0,T)=\left( \frac{1}{2\pi  T} \right)^{1/2} \exp \left[\frac{-v_0^2}{2T} \right].
\ee
As discussed earlier, we can effectively treat the particles as non-interacting and the problem reduces to a single particle problem~\cite{Roy2013}. We then consider  a single particle starting from an initial position $x_0$ with initial velocity $v_0$.  The final position, $x_t$, of the particle, taking all possible collisions into account, can take the following forms:
\begin{align}
x_t =\begin{cases}
 x_0+v_0t-2nL, &{\text {~~if~~} } v_0>0,v_t>0, \\ 
 x_0+v_0t+2nL, &{\text {~~if~~} } v_0<0,v_t<0, \\
 -x_0-v_0t+2nL,  &{\text {~~if~~} } v_0>0,v_t<0, \\ 
-x_0-v_0t-2nL,  &{\text {~~if~~} } v_0<0,v_t>0, 
\label{eq:xt} 
\end{cases}
\end{align}
where $v_t$, the velocity at time $t$, can take either values $\pm v_0$  and $n=0, 1, 2, 3 ...,$ is the number of collision(s) that the particle has with both the boundaries. The distribution $F(x,v,t)$ is then obtained by averaging over the initial position [uniform in $(0,L/2)$] and velocity [drawn from $g_{\rm eq}(v_0,T)$]:
\be
\label{f_eq0}
F(x,v,t)=N\langle \delta(x-x_t) \delta(v-v_t) \rangle.
\ee
\begin{widetext}
Using Eq.~\eqref{eq:xt} we get:
\begin{align} 
&F(x,v,t) = \frac{2N}{L} \int_0^{L/2} dx_0 \int_{-\infty}^{\infty}dv_0 g_{\rm eq}(v_0) \sum_{n=-\infty}^{\infty} \left[ \delta(x-x_0-v_0t+2nL) 
 \delta(v-v_0) + \delta(x+x_0+v_0t-2nL) \delta(v+v_0) \right], \label{f_eq1} \\ 
\label{f_eq2}
&= {2\rho_0} \frac{\exp (-v^2/2T)}{\sqrt{2\pi T}}  \sum_{n=-\infty}^{\infty} [ \Theta(x-vt-2nL+L/2) - \Theta(x-vt+2nL-L/2) ],
\end{align}
where $\rho_0=N/L$.  Now we calculate the three mean fields by performing the integrals in  Eqs.~(\ref{eq_d_field}), (\ref{eq_p_field}), and (\ref{eq_e_field}) to get
\begin{align} \nonumber
\bar{\rho}(x,t)&  =  \int_{-\infty}^{\infty}dv F(x,v,t), \\ \nonumber
&= {2\rho_0} \frac{1}{\sqrt{2\pi T}} \sum_{n=-\infty}^{\infty} \int_0^{L/2} dx_0 \int_{-\infty}^{\infty}dv~ e^{-v^2/2T} \left[ \delta(x-x_0-vt+2nL) 
+ \delta(x+x_0-vt-2nL) \right], \\ \nonumber
&= 2\rho_0 \frac{1}{\sqrt{2\pi T}} \sum_{n=-\infty}^{\infty} \int_0^{L/2} dx_0 \frac{1}{t} \left[ \exp \left\{ \frac{-(2nL+x-x_0)^2}{2Tt^2} \right\} 
+ \exp \left\{ \frac{-(2nL-x-x_0)^2}{2Tt^2} \right\} \right], \\
\label{eq_density_field}
&= \rho_0 \sum_{n=-\infty}^{\infty} \left[\text{erf}\left( \frac{L/2-2nL-x}{\sqrt{2T}t} \right) + \text{erf}\left( \frac{L/2-2nL+x}{\sqrt{2T}t} \right) \right].
\end{align}
Using the Poisson resummation formula, this can  be rewritten in the alternative series form: 
\begin{align}
\label{eq_density_field2}
\bar{\rho}(x,t)= \rho_0 + 4\rho_0 \sum_{k=1}^{\infty} \frac{1}{k\pi} \sin \left( \frac{k\pi}{2} \right) \cos \left( \frac{k\pi x}{L} \right) \exp \left( \frac{-k^2\pi^2Tt^2}{2L^2} \right).
\end{align}
Following a similar approach we obtain the following expressions for the mean momentum and energy:
\begin{align} \nonumber
\bar{p}(x,t)& = \int_{-\infty}^{\infty}dv~v F(x,v,t), \\
\label{eq_momentum_field}
&= \rho_0 \sqrt \frac{2T}{\pi} \sum_{n=-\infty}^{\infty} \left[ \exp \left\{- \frac{(2nL-L/2+x)^2}{2Tt^2} \right\}  
 - \exp \left\{- \frac{(2nL-L/2-x)^2}{2Tt^2} \right\} \right], \\
\label{eq_momentum_field2}
&= \frac{4\rho_0Tt}{L} \sum_{k=1}^{\infty} \sin \left( \frac{k\pi}{2} \right) \sin \left( \frac{k\pi x}{L} \right) \exp \left( \frac{-k^2\pi^2Tt^2}{2L^2} \right), \\ \nonumber
\bar{e}(x,t)& = \frac{1}{2} \int_{-\infty}^{\infty}dv~v^2 F(x,v,t), \\ \nn
&= \rho_0 \frac{2T}{\sqrt{\pi}} \sum_{n=-\infty}^{\infty} \left[
-\frac{\sqrt{\pi}}{4} \text{erf}\left( \frac{2nL-L/2+x}{\sqrt{2T}t} \right) + \frac{2nL-L/2+x}{2\sqrt{2T}t} \exp \left\{\frac{-(2nL-L/2+x)^2}{2Tt^2}\right\} \right. \\
\label{eq_energy_field}
& \left. -\frac{\sqrt{\pi}}{4} \text{erf}\left( \frac{2nL-L/2-x}{\sqrt{2T}t} \right) + \frac{2nL-L/2-x}{2\sqrt{2T}t} \exp \left\{\frac{-(2nL-L/2-x)^2}{2Tt^2}\right\} \right]. \\
\label{eq_energy_field2}
&= \frac{\rho_0T}{2} + 2\rho_0T \sum_{k=1}^{\infty} \frac{1}{k\pi} \left(1- \frac{k^2\pi^2Tt^2}{L^2} \right) \sin \left( \frac{k\pi}{2} \right) \cos \left( \frac{k\pi x}{L} \right) \exp \left( \frac{-k^2\pi^2Tt^2}{2L^2} \right).
\end{align}
From Eqs.~(\ref{eq_density_field2},~\ref{eq_momentum_field2},~\ref{eq_energy_field2}) one can easily see that the approach to equilibrium has the long-time form $e^{-a t^2}$ with $a=T \pi^2/(2 L^2)$.
\end{widetext}

\section{Numerical methods}
\label{app:numerics}

Here we describe the details of the simulation procedure. We consider a box $(0,L)$ with particles initially distributed uniformly in the left half $(0,L/2)$ of the box. The velocity of each particle is chosen independently from a specified distribution.  Since the particles are non-interacting, their velocities  change only during collisions with the walls at $x=0$ and $x=L$.  The position of each particle is updated according to Eqs.~\eqref{eq:xt} which incorporates the effect of reflecting boundaries. With the  system evolving in time with this dynamics,  we compute the required observables.

\end{document}